\documentclass[aps,prd,twocolumn,showpacs,showkeys,superscriptaddress]{revtex4-1}
\usepackage{amsmath,amssymb}
\usepackage{graphicx}

\newcommand{\bastar}{\begin{eqnarray*}}
\newcommand{\eastar}{\end{eqnarray*}}
\newskip\humongous \humongous=0pt plus 1000pt minus 1000pt

\newif\ifdtup

\relax

\newcommand{\bea}{\begin{eqnarray}}
\newcommand{\eea}{\end{eqnarray}}

\newcommand{\nn}{\nonumber}
\newcommand{\pro}{\partial}

\newcommand{\mn}{{\mu\nu}}
\begin{document}
\title{Cosmological Production of Electroweak Monopole}
\bigskip
\author{Y. M. Cho}
\email{ymcho7@konkuk.ac.kr}
\affiliation{Administration Building 310-4, 
Konkuk University, Seoul 143-701, Korea}
\affiliation{School of Physics and Astronomy, 
Seoul National University,
Seoul 151-742, Korea}
\author{Kyoungtae Kimm}
\affiliation{Faculty of Liberal Education,
Seoul National University, Seoul 151-747, Korea}
\author{Seunghun Oh}
\affiliation{Institute of Basic Science, College of Natural 
Sciences, Konkuk University, Seoul 143-701, Korea}
\author{J. H. Yoon}
\affiliation{Department of Physics, College of Natural 
Sciences, Konkuk University, Seoul 143-701, Korea}

\begin{abstract}
We discuss the cosmological production and the successive
evolution of the electroweak monopole in the standard
model, and estimate the remnant monopole density at present 
universe. We confirm that, although the electroweak phase 
transition is of the first order, it is very mildly first 
order. So, the monopole production arises from the thermal 
fluctuations of the Higgs field after the phase transition, 
not the vacuum bubble collisions during the phase transition. 
Moreover, while the monopoles are produced copiously around 
the Ginzburg temperature $T_G\simeq 59.6~{\rm TeV}$, most 
of them are annihilated as soon as created. This annihilation 
process continues very long, untill the temperature cools 
down to about 29.5 MeV. As the result the remnant monopole 
density in the present universe becomes very small, of 
$10^{-11}$ of the critical density, too small to affect 
the standard cosmology and too small comprise a major 
component of dark matter. We discuss the physical implications 
of our results on the ongoing monopole detection experiments, 
in particular on MoEDAL, IceCube, ANTARES, and Auger.         
\end{abstract}


\maketitle

\section{Introduction}

Ever since Dirac has proposed the Dirac monopole 
generalizing the Maxwell's theory, the monopole 
has become an obsession theoretically as well as
experimentally \cite{dirac}. After the Dirac monopole 
we have had the Wu-Yang monopole \cite{wu}, 
the 't Hooft-Polyakov monopole \cite{thooft}, 
the grand unification (Dokos-Tomaras) monopole 
\cite{dokos}, and the electroweak (Cho-Maison) 
monopole \cite{plb97,yang,epjc15,ellis,plb16}. But 
none of them except for the electroweak monopole 
might become realistic enough to be discovered. 

Indeed the Dirac monopole in electrodynamics should 
transform to the electroweak monopole after 
the unification of the electromagnetic and weak 
interactions, and the Wu-Yang monopole in QCD is 
supposed to make the monopole condensation to 
confine the color. Moreover, the 't Hooft-Polyakov 
monopole exists in an hypothetical theory, and 
the grand unification monopole which could have 
been amply produced at the grand unification scale 
in the early universe probably has become completely 
irrelevant at present universe after the inflation. 

This makes the electroweak monopole the only realistic 
monopole we could ever hope to detect. This has made 
the experimental confirmation of the electroweak 
monopole one of the most urgent issues in the standard 
model, after the discovery of the Higgs particle at 
LHC. In fact the newest MoEDAL (``the magnificent 
seventh") detector at LHC is actively searching for 
the monopole \cite{medal1,medal2}.

On the other hand, MoEDAL may have no chance to 
detect the electroweak monopole if the mass becomes 
larger than 6.5 TeV, because the present 13 TeV LHC 
is expected to produce the monopole pair only when 
the mass is smaller than 6.5 TeV. In this case we 
may have to search for the remnant monopole in 
the present universe which survived from the early 
universe, or else wait for the next LHC upgrading. 
So it is wise to adopt two track strategy to detect 
the monopole, the LHC-produced monopoles and the remnant 
monopoles in the present universe. In this sense it 
is encouraging that IceCube and similar detectors, 
e.g., ANTARES, Auger, are aiming to detect the remnant 
monopoles \cite{icecube,antares,auger}.   

To detect the remnant monopoles, however, it is important 
for us to know the monopole density at present universe. 
There have been discussions on the impact of the monopoles 
in cosmology \cite{kibble,pres,guth,zurek}, but most 
of the discussions have been on the grand unification 
monopoles. The general consensus is that the grand 
unification monopoles would have overclosed the universe 
without inflation, but inflation might have completely 
diluted them in such a way that the grand unification 
monopole could have no visible impact on the present 
universe \cite{infl}.

For the electroweak monopole, however, the situation 
has  not been so clear \cite{kriz,zel,and,koba}. There 
have been claims that the electroweak monopoles could 
also overclose the universe \cite{zel}, but we need 
a more accurate discussion on this issue. The purpose 
of this paper is to discuss the monopole production after 
the electroweak phase transition and the successive 
cosmological evolution of the monopoles, to predict 
the remnant monopole density at present universe and 
to discuss the cosmological implications of the electroweak 
monopole.

Our analysis confirms that the electroweak phase transition 
is of the first order, but the energy barrier at the critical 
temperature is negligibly small that it can be treated 
as the second order for all practical purposes. So we argue 
that the monopole production comes from the thermal 
fluctuations of the Higgs field after the phase transition, 
not the vacuum bubble collosions during the phase 
transition. As importantly, we show that although 
the electroweak monopoles are produced copiously around 
the Ginzburg temperature $T_G\simeq 60~{\rm GeV}$, 
they are annihilated as soon as created. And this 
annihilation process continues very long time, till 
the universe cools down to around 29.5 MeV. This is 
basically because the monopole-antimonopole attraction 
is much bigger, $1/\alpha$-times bigger, than the 
electromagnetic interaction. 

As the result the density of the electroweak monopoles 
at present universe turns out to be very small, about 
$10^{-11}$ of the critical density. This assures that, 
unlike the grand unification monopole, the electroweak 
monopole does not cause any  problem in cosmology. 
Moreover, this tells that there is no possibility that 
the electroweak monopoles could become the dark 
matter of the universe. 

Nevertheless we find that there are enough remaining 
monopoles in the present universe which we could detect 
without much difficulty. We estimate the monopole number 
density at present universe to be roughly $6.1 \times 10^{-22} 
/ {\rm cm^3}$. Intuitively, this means that there are roughly 
$6.6 \times 10^7$ monopoles per unit volume of the earth
in the universe.

Although the electroweak monopole does not alter the standard 
cosmology, it could play important roles in cosmology. First,
as the heaviest stable particle the remnant monopoles could 
generate the density perturbation and become the seeds of 
the large scale structures of the universe. Second, accelerated 
by the intergalactic magnetic field, they become the natural 
source of the ultra-relativistic cosmic rays. Third, they could 
induce the electroweak baryogenesis.   
   
The paper is organized as follows. In Section II we review 
the fundamental properties of the electroweak monopole 
for later purpose. In Section III we discuss the electroweak 
phase transition in detail. In Section IV we discuss 
the monopole production after the phase transition in
detail, both in the first order phase transition and the second 
order approximation. In Section V we discuss the evolution 
of the monopoles and estimate the remnant monopole 
density at present universe. In Section VI we compare 
our result with the Parker bound on average intergalactic 
magnetic field in the universe. In the last section we discuss 
the physical implications of our results, in particular on 
the ongoing MoEDAL, IceCube, and similar monopole 
detection experiments.   

\section{Electroweak Monopole: A Review}

Before we estimate the electroweak monopole density 
at present universe, we need to clarify the existing 
confusions and misunderstandings on the monopole. 
For this reason we briefly review the electroweak 
monopole in the standard model first. 

\begin{figure}
\includegraphics[width=8cm, height=4.5cm]{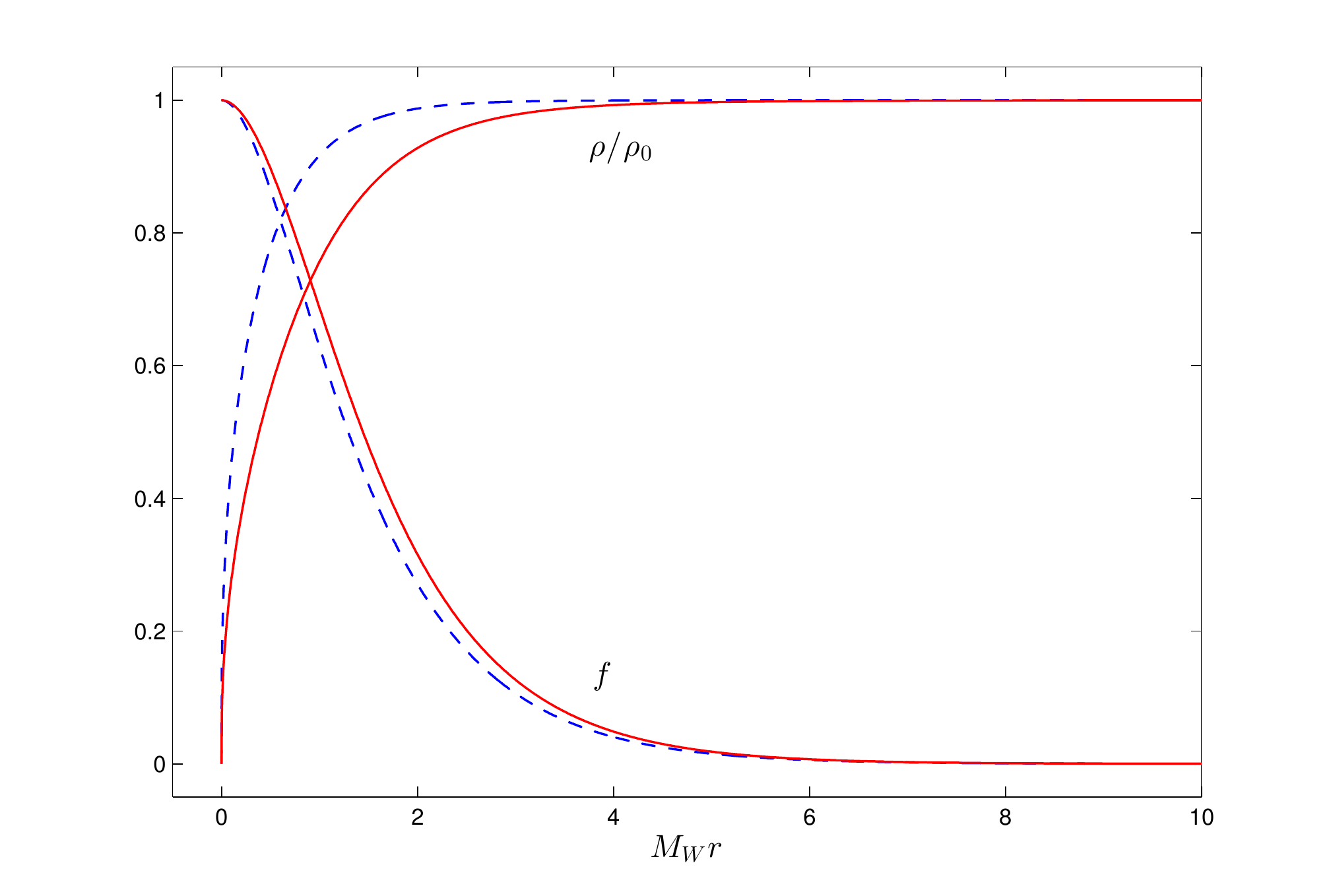}
\caption{\label{fcmono} The finite energy electroweak 
monopole solution. The solid line (red) represents 
the regularized monopole and the dotted (blue) line 
represents the singular electroweak monopole.}
\end{figure}

Consider the Weinberg-Salam model,
\begin{gather}
{\cal L}= -|{\cal D} _\mu \phi|^2
-\frac{\lambda}{2} \Big(\phi^\dagger \phi 
-\frac{\mu^2}{\lambda} \Big)^2 
-\frac{1}{4} \vec F_\mn^2 -\frac{1}{4} G_\mn^2,  \nn\\
{\cal D}_{\mu} \phi=\big(\partial_{\mu} 
-i\frac{g}{2} \vec \tau \cdot \vec A_{\mu} 
- i\frac{g'}{2}B_{\mu}\big) \phi,
\label{wsl1}
\end{gather}
where $\phi$ is the Higgs doublet, $\vec A_\mu$ and 
$B_\mu$ are the SU(2) and U(1) gauge potentials. Introducing 
the Higgs field $\rho$ and the $CP^1$ field $\xi$ by
\bea
\phi = \dfrac{1}{\sqrt{2}}\rho~\xi,
~~~(\xi^{\dagger} \xi = 1),
\eea
we have
\begin{gather}
{\cal L}=-\frac{1}{2}{(\partial_{\mu}\rho)}^2
- \frac{\rho^2}{2} {|{\cal D}_{\mu} \xi |}^2
-\frac{\lambda}{8}\big(\rho^2-\rho_0^2\big)^2 \nn\\
-\frac{1}{4}{\vec F}_\mn^2 -\frac{1}{4} G_\mn^2,
~~~~(\rho_0^2=\frac{2\mu^2}{\lambda}).
\label{wsl2}
\end{gather}
With the ansatz
\begin{gather}
\rho =\rho(r),  
~~~\xi=i\left(\begin{array}{cc} \sin (\theta/2)~e^{-i\varphi}\\
- \cos(\theta/2) \end{array} \right),   \nn\\
\vec A_{\mu}= \frac{1}{g} A(r)\partial_{\mu}t~\hat r
+\frac{1}{g}(f(r)-1)~\hat r \times \pro_{\mu} \hat r, \nn\\
B_{\mu} =\frac{1}{g'} B(r) \partial_{\mu}t 
-\frac{1}{g'}(1-\cos\theta) \partial_{\mu} \varphi,
\label{ans}
\end{gather}
we have the dyon solution of the standard model 
dressed by the W-boson, Z-boson, and Higgs field, 
which becomes the monopole solution with $A=B=0$. 
The singular monopole solution is shown in 
Fig. \ref{fcmono} \cite{plb97,yang,epjc15}. 

The first and (most serious) misunderstanding on
the electroweak monopole is the existence. This, of 
course, is the fundamental issue. It has been asserted 
that the vacuum of the Weinberg-Salam model does 
not allow the monopole topology \cite{nogo}. If so, 
there would be no phase transition which could produce 
the monopole, and thus no cosmic production of 
the monopole. 

Actually there are two questions here. As we know, 
the Dirac monopole in electrodynamics is optional, 
not a necessity, so that it does not have to exist. 
This is because the electromagnetic U(1) allows 
the monopole only when it becomes non-trivial. So 
the first question is whether the standard model 
could admit the monopole topology or not. The second 
question is, if the monopole topology is consistent with 
the standard model, is the monopole optional or not. 

Obviously the existence of the monopole assures that 
the standard model could be made to admit the monopole 
topology \cite{plb97,yang,epjc15}. So the remaining 
question is if this monopole topology is optional or 
not. It is not optional. The electroweak monopole 
must exist, if the standard model is correct.
 
To see this notice that in the ansatz (\ref{ans}) 
the hypercharge U(1) is made to be non-trivial, which 
was why the electroweak monopole could exist. So 
we have to find the reason why the hypercharge U(1) 
in the standard model must be non-trivial. This follows 
from two facts. First, the electromagnetic U(1) in 
the standard model is given by the linear combination 
of the U(1) subgroup of SU(2) and the hypercharge U(1).
Second, the U(1) subgroup of SU(2) is non-trivial. In this 
case the mathematical consistency requires the hypercharge 
U(1), and consequently the electromagnetic U(1), to be 
non-trivial. 

But, of course, this has to be confirmed by experiment. 
This makes the discovery of the monopole, not the Higgs 
particle, the final (and topological) test of the standard 
model. This is why the electroweak monopole is so 
important.

The second confusion is the assertion that, even if 
it exists, there would be no fundamental difference 
between the Dirac monopole and the electroweak 
monopole. So the important question here is whether 
there is any characteristic feature of the electroweak 
monopole which is different from the Dirac monopole. 
We have to know this to tell the monopole, when 
discovered, is the Dirac monopole or the electroweak 
monopole. The answer is yes, there is an unmistakable 
difference. The electroweak monopole carries 
the magnetic charge twice bigger than the Dirac 
monopole \cite{plb97,epjc15}. 

The reason is simple. It is well known that the magnetic 
charge of Dirac monopole must be a multiple of $2\pi/e$. 
This is because the period of the electromagnetic U(1) is 
$2\pi$. However, in the course of the electroweak 
unification the period of electromagnetic U(1) becomes 
$4\pi$, because the period of the U(1) subgroup of SU(2) 
becomes $4\pi$. Consequently the magnetic charge 
$g_m$ of the electroweak monopole must be a multiple 
of $4\pi/e$. 

The third misunderstanding is that we can not estimate 
the mass of the electroweak monopole, because 
the electroweak monopole has the point singularity at 
the origin which makes the energy infinite. Obviously 
the monopole mass is the most important piece of 
information from the experimental point of view. 
There was no way to predict the mass of the Dirac 
monopole theoretically, and this has made the monopole 
detection a blind search in the dark room without any 
theoretical lead. 

But this assertion is not true, either. Since the mass 
of the electroweak monopole is a crucial information 
for us to estimate the monopole density in the universe, 
it is worth to discuss this issue in more detail. As 
a hybrid between the Dirac and 't Hooft-Polyakov 
monopoles, the electroweak monopole does have 
a singularity at the origin which makes the energy 
divergent. Nevertheless we could easily guess 
the mass to be of the order of 10 TeV, roughly 
$4\pi/e^2$ times the W-boson mass. This is because 
the monopole mass essentially comes from the same 
Higgs mechanism which generates the W-boson mass, 
except that the monopole potential couples to the Higgs 
multiplet magnetically, not electrically, with the strength 
$4\pi/e$ \cite{plb97,epjc15}.  

Another way to estimate the mass is to regularize 
the monopole to make the energy finite \cite{epjc15}. 
To do this let us modify (\ref{wsl2}) to the following 
effective Lagrangian with a non-trivial permittivity 
$\epsilon$ for the hypercharge U(1) gauge field 
\begin{gather}
{\cal L}_{eff}=-\frac{1}{2}{(\partial_{\mu}\rho)}^2
- \frac{\rho^2}{2} {|{\cal D}_{\mu} \xi |}^2
-\frac{\lambda}{8}\big(\rho^2-\rho_0^2\big)^2 \nn\\
-\frac{1}{4}{\vec F}_\mn^2 
-\frac{\epsilon(\rho)}{4} G_\mn^2.
\label{effl}
\end{gather}
The effective Lagrangian still retains the SU(2) $\times$ 
U(1) gauge symmetry. Moreover, when $\epsilon =1$, 
the Lagrangian reproduces the standard model. So this 
modification affects only the short distance behavior, 
if $\epsilon$ approaches to the unit asymptotically.

To see how $\epsilon$ mimics the quantum correction 
and regularize the monopole, notice that the rescaling 
of $B_\mu$ to $B_\mu/g'$ changes $g'$ to 
$g' /\sqrt{\epsilon}$. So $\epsilon$ effectively changes 
the hypercharge U(1) gauge coupling $g'$ to the ``running" 
coupling  $\bar g'=g' /\sqrt{\epsilon}$. This means that, 
by making $\bar g'$ infinite (requiring $\epsilon$ 
vanishing) at the origin, we can remove the singularity 
of the monopole. Moreover, with $\epsilon(\infty)=1$, 
we can reproduce the singular monopole asymptotically. 
In fact, with $\epsilon=(\rho/\rho_0)^8$, we have 
the regularized monopole solution which has the energy 
around 7 TeV \cite{epjc15,plb16}. This is shown in 
Fig. \ref{fcmono}. Notice that asymptotically 
the regularized monopole looks almost identical to 
the singular monopole. 

The regularized monopole energy, of course, depends on 
the functional form of $\epsilon$, so that we could change 
the energy changing $\epsilon$. But this also affects other 
things, for example the Higgs to two photon decay rate. 
And recently Ellis and collaborators noticed that, choosing 
a more realistic $\epsilon$ which can reproduce the experimental 
value of the Higgs to two photon decay rate, they could
reduce the monopole energy less than 5.5 TeV. 

Furthermore, we can show that the gravitational interaction 
does not change the monopole mass much. This is important, 
because the gravitational interaction could turn the monopole
to a black hole, in which case the monopole mass could 
not be predicted. Coupling the effective Lagrangian (\ref{effl}) 
to the Einstein's gravity, we can obtain a family of globally 
regular gravitating electroweak monopole solutions whose 
ADM mass are almost the same as the monopole energy 
without the gravitational interaction \cite{plb16}. Moreover, 
we can show that these gravitating monopoles turn to 
the magnetically charged black holes only when the Higgs 
vacuum value approaches to the Planck mass. 

From this we can conclude that the mass of the electroweak 
monopole can be predicted, and it must be around 4 to 
10 TeV. This is encouraging and at the same time tantalizing, 
because this tells that LHC could not produce the monopoles 
if the mass is larger than 6.5 TeV. Under this circumstance, it 
is important for us to estimate the monopole density at present 
universe. For this purpose we discuss the electroweak phase 
transition first.

\section{Electroweak Phase Transition}

It is generally believed that the monopole production 
in the early universe comes from the phase transition. 
At Planck time ($t\simeq 10^{-44}~s$) all interactions are 
supposed to be unified symmetrically in the unified group G,
and the universe is in the normal phase. But as the universe 
cools down, it has at least two distinct stages of symmetry 
breaking. At the grand unification scale around $10^{15}$ 
GeV, G breaks down to the unbroken subgroup H made of 
the color SU(3), the weak SU(2), and the hypercharge U(1). 
At the much lower electroweak scale of $10^{2}$ GeV, this 
symmetry H breaks down further to the color SU(3) and 
the electromagnetic U(1). 

This of course is the simplest possible scenario, but is supported 
by the renormalization group calculation which shows that all
three coupling constants associated to SU(3), SU(2), and U(1)
become equal to the value of $\alpha=g^2/4\pi\simeq 1/50$.
 
And these symmetry breakings induced spontaneously 
by the Higgs mechanism are expected to induce the change 
of topology and generate topological objects when 
the manifold M=G/H of the degenerate vacuua determined 
by the Higgs field has non-trivial homotopy group, where 
H is the unbroken subgroup of G. For example, we expect 
the domain walls when $\pi_0(M)$ is non-trivial, the strings 
when $\pi_1(M)$ is non-trivial, and the monopoles when 
$\pi_2(M)$ is non-trivial.

So far much of the attention has been payed to the grand 
unification symmetry breaking and the resulting grand 
unification monopole production. It has been argued that 
at this stage (around $t \simeq 10^{-37}~s$) the massive 
grand unification monopoles should have been amply 
produced, so much so that their mass density would 
exceed that of all other matters by many orders of 
magnitude \cite{pres}. This was one of the reasons to justify 
the cosmic inflation which could dilute the monopoles 
completely \cite{infl}. 

What we are concerned is the later stage of the phase 
transition at the electroweak scale, at around 
$t\simeq 10^{-11}~s$. To study the behavior of 
the electroweak theory at this stage, we have to 
compute the temperature-dependent one-loop 
correction to the Higgs potential. Fortunately 
this has already been done, and the effective potential 
can be written as \cite{kriz,and},
\begin{gather}
V_T(\rho) =V_0(\rho) -\frac{C_1}{12\pi} \rho^3~T
+\frac{C_2}{2} \rho^2~T^2 
-\frac{\pi^2}{90} N T^4+\delta V_T,  \nn\\
V_0(\rho)=\frac{\lambda}{8}(\rho^2-\rho_0^2)^2 ,  \nn\\
C_1=\frac{6 M_W^3 + 3 M_Z^3}{\rho_0^3}\simeq 0.36,   \nn\\
C_2=\frac{4M_W^2 +2 M_Z^2 +M_H^2+4m_t^2}{8\rho_0^2} 
\simeq 0.36,   
\label{epot}
\end{gather}
where $V_0$ is the zero-temperature Higgs potential, 
$N$ is the total number of distinct helicity states of 
the particles with mass smaller than $T$ (counting 
fermions with the factor 7/8), $C_1$ and $C_2$ are 
the constants fixed by the weak boson and heavy quark 
masses, and $\delta V_T$ is the slow-varying logarithmic 
corrections and the lighter quark contributions given by
\begin{gather}
\delta V_T=-\frac{M_H^4}{256\pi^2} \log \frac{M_H^2 (3\rho^2 
-\rho_0^2)}{2a \rho_0^2 T^2}+ \frac{M_H^3 T \sqrt{2(3\rho^2 
- \rho_0^2)}}{ 48 \pi \rho_0}  \nn \\ 
+\frac{3M_H^4 \rho^2}{128 \pi^2\rho_0^2}\log \frac{M_H^2(3\rho^2
-\rho_0^2)}{2a \rho_0^2 T^2} -\frac{M_H^3 T \rho^2\sqrt{2(3\rho^2 
- \rho_0^2)}}{16\pi \rho_0^3}   \nn \\
-\frac{3\rho^4}{64\pi^2 \rho_0^4} \Big[ 2 M_W^4 
\log\frac{M_W^2 \rho^2 }{a \rho_0^2 T^2} 
+M_Z^4\log\frac{M_Z^2 \rho^2}{a \rho_0^2 T^2}  \nn \\
+\frac{3M_H^4}{4}\log\frac{M_H^2(3\rho^2-\rho_0^2)}{2a \rho_0^2 T^2} 
-4 m_t^4 \log\frac{m_t^2 \rho^2}{a \rho_0^2 T^2} \Big],   \nn\\
a = 16\pi^2 \exp(3/2-2\gamma_E),
\end{gather} 
where $\gamma_E$ is the Euler-Mascheroni constant.
But in the following we will neglect 
this term for simplicity. Experimentally we have $\lambda\simeq$ 
0.258, $\rho_0\simeq$ 246 GeV, $M_W\simeq$ 80.4 GeV, 
$M_Z\simeq$ 91.2 GeV, $M_H\simeq$ 125.7 GeV, and 
$m_t \simeq$ 173.2 GeV.

\begin{figure}
\includegraphics[height=4.5cm, width=7cm]{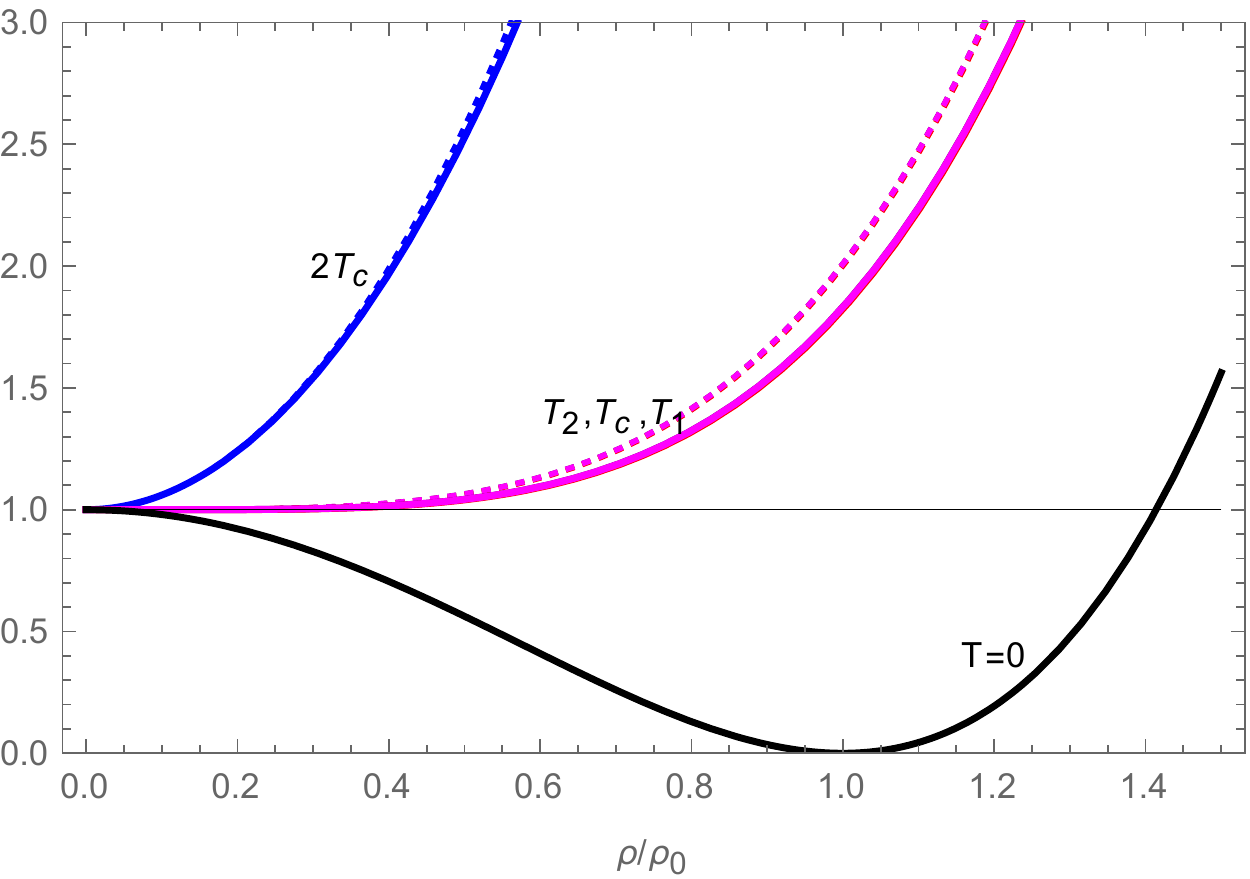}
\caption{\label{tpot1} The effective thermal potential 
$V_{eff}=V_T-(\pi^2 N/90)~NT^4$ (the solid lines) for 
$T=2 T_c, T_2, T_c, T_1,$ and $0$. For comparison 
we plot the approximated potential shown in Eq. (12) in 
dashed lines. Here the unit of $V_{eff}$ is chosen to be 
$V_0=(\lambda/8) \rho_0^4=1$, and the horizontal line 
represents the $V_{eff}=V_0$. Notice that $V_T$ is almost 
indistinguishable at $T_2, T_c$, and $T_1$.}
\end{figure} 

The effective potential (with $\delta V_T=0$) has three local 
extrema at 
\begin{gather}
\rho_s=0,   \nn\\
\rho_{\pm}(T)=\frac{C_1}{4\pi \lambda}~T 
\pm \sqrt{ \Big(\frac{C_1}{4\pi \lambda} \Big)^2 ~T^2
+\rho_0^2 -\frac{2C_2}{\lambda}~T^2}.
\label{rext}
\end{gather}
The first extremum $\rho_s=0$ represents the Higgs vacuum 
of the symmetric (unbroken) phase, and the second extremum 
$\rho_-(T)$ represents the local maximum, and the third 
extremum $\rho_+(T)$ represent the local minimum Higgs 
vacuum of the broken phase. But notice that these two 
extrema $\rho_{\pm}$ appear only when $T$ becomes smaller 
than 
$T_2$
\begin{gather}
T_2=\frac{T_1}{\sqrt{1-\alpha^2}}\simeq 146.7~{\rm GeV},  \nn\\
T_1=\sqrt{\frac{\lambda}{2C_2}}~\rho_0 
\simeq  146.4~{\rm GeV},   \nn\\
\alpha=\frac{C_1}{4\pi \sqrt{2\lambda C_2}} \simeq 0.0662.
\end{gather}
So above this temperature only $\rho_s=0$ becomes the true 
vacuum of the effective potential, and the electroweak symmetry 
remains unbroken. 

At $T=T_2$ we have 
\begin{gather}
\rho_-=\rho_+=(C_1/4\pi \lambda)~T_2
\simeq 16.3~{\rm GeV},
\end{gather} 
but as temperature cools down below $T_2$ we have two local 
minima at $\rho_s$ and $\rho_+$ with $V_T(0)< V_T(\rho_+)$,
until $T$ reaches the critical temperature $T_c$ where 
$V_T(0)$ becomes equal to $V_T(\rho_+)$,
\begin{gather}
T_c=\frac{T_1}{\sqrt{1- 8\alpha^2/9}} 
\simeq 146.6 ~{\rm GeV},   \nn\\
\rho_+(T_c)=\frac{C_1}{3\pi \lambda}~T_c
\simeq 21.8~{\rm GeV}.
\label{ctemp}
\end{gather}
So $\rho_s=0$ remains the minimum of the effective potential
for $T>T_c$. Notice that $T_c/\rho_0\simeq 0.6$ but 
$\rho_+/\rho_0 \simeq 0.09$. 

\begin{figure}
\includegraphics[height=4.5cm, width=7cm]{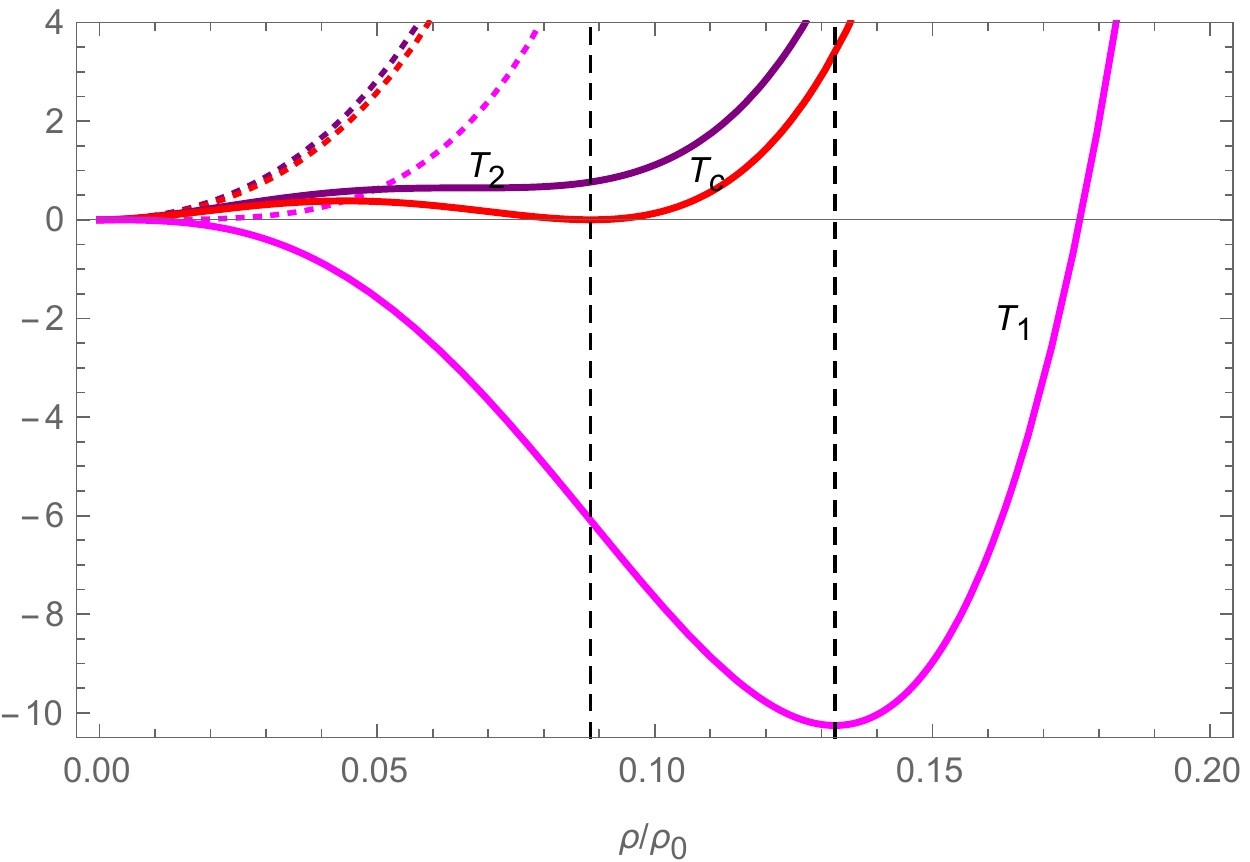}
\caption{\label{tpot2} The amplified effective thermal 
potential for $T=T_2, T_c$, and $T_1$. Here the vertical 
axis represents $V_{eff}-V_0$ in the unit $10^{-5}\times V_0$, 
and the vertical dashed lines indicate the positions of 
the local minima of the potentials. Here again the dashed 
lines represents the approximated potential (12).}
\end{figure}

Below this critical temperature $\rho_+$ becomes the true 
minimum of the effective potential, but $\rho_s=0$ remains 
a local (unstable) minimum till the temperature reaches 
$T_1$. But at $T=T_c$ the new vacuum bubbles start to 
nucleate at $\rho=\rho_+$, which takes over the unstable 
vacuum $\rho_s=0$ completely at $T=T_1$ when $\rho_+(T_1)$ 
becomes around 32.6 GeV. From this point $\rho_+$ becomes 
the only (true) minimum, which approachs to the well-known 
Higgs vacuum $\rho_0$ at zero temperature. 

The temperature dependence of the effective potential is 
schematically shown in Fig. \ref{tpot1}. But since $V_T$ 
is almost indistinguishable at $T_2, T_c$, and $T_1$,
we have amplified it in Fig. \ref{tpot2}. Notice that 
around the critical temperature $T\simeq T_c$ we have 
$\rho_{\pm}(T) \ll T$. This assures that the high temperature 
approximation of the effective potential (\ref{epot}) is 
trustable.  

Fig. \ref{tpot2} tells that the phase transition from 
the local minimum from $\rho_s$ to the true minimum 
$\rho_+$ after $T=T_c$ is classically forbidden till 
$T$ reaches $T_1$, because the two minima are separated 
by an energy barrier. So during this period the transition 
must take place slowly through the quantum tunneling. 
Below this temperature the energy barrier disappears 
and we have free (fast) phase transition which generates 
a large latent heat. This means that the electroweak 
phase transition is of the first order.

In reality, however, Fig. \ref{tpot1} shows that the energy 
barrier is very small, which confirms that the phase 
transition is almost second order \cite{and}. To see 
this notice first that from (\ref{ctemp}) we have 
$\rho_+(T_c)/T_c \simeq 0.148$, which tells that 
the two degenerate vacua at $T=T_c$ are very close. 
And we can easily show that the height of the barrier 
between these vacua is extremely small,
\begin{gather}
\frac{V_{T_c}(\rho)|_{max}-V_0}{V_0}
\simeq 3.83\times 10^{-6}, 
~~~V_0=\frac{\lambda}{8} \rho_0^4.
\end{gather} 
Moreover, the barrier lasts only for short period since 
the temperature difference from $T_c$ to $T_1$ is very 
small, $\delta =(T_c-T_1)/T_c \simeq 0.001$. From this 
we conclude that the electroweak phase transition is 
very mildly first order, in fact almost the second order.

Notice that it is the second term in (\ref{epot}) linear 
in $T$ which makes the electroweak phase transition first 
order. But this term does not change the effective 
potential much since it has a small coefficient compared 
to the third term (i.e., $C_1/12\pi \ll  C_2/2$), and thus 
can be neglected. Neglecting this term we can approximate 
the effective potential to
\begin{gather}
V_T(\rho) \simeq V_0(\rho) +\frac{C_2}{2} \rho^2~T^2
-\frac{\pi^2}{90} N T^4.   
\label{epot2}   
\end{gather}
In this approximation we have 
$T_c=\sqrt{\lambda/2C_2} \rho_0=T_2=T_1=146.4$ GeV, 
so that $T_c$, $T_1$, and 
$T_2$ of the first order phase transition all become 
the same. 

The effective potential (\ref{epot2}) has only two minima, 
$\rho_s$ for $T>T_c$ and $\rho_+$ for $T<T_c$,
\begin{gather}
\rho_s=0,   \nn\\
\rho_+(T)=\sqrt{1-\Big(\frac{T}{T_c}\Big)^2} \rho_0, 
\end{gather}
so that the phase transition becomes exactly the second 
order. For comparison this potential is plotted in 
Fig.\ref{tpot1} and Fig. \ref{tpot2} in dashed lines. 

However, there is one big difference between two effective 
potentials (\ref{epot}) and (\ref{epot2}). In the first 
order phase transition the Higgs mass remains non-vanishing 
during the phase transition. From (\ref{epot}) we have
\begin{gather}
\bar M_H^2=\frac{d^2 V_{eff}}{d\rho^2} \Big|_{\rho_{min}}  \nn\\
=\left\{\begin{array}{ll} 
\big[(T/T_1)^2-1 \big] M_H^2/2,&~~~T \ge T_c,  \\
\big[(\rho_+/\rho_0)^2 +1-(T/T_1)^2 \big] M_H^2/2,&~~~T<T_c, 
\end{array} \right.
\end{gather}
where $\bar M_H$ is the temperature-dependent Higgs mass. 
So $\bar{M}_H$ acquires its minimum value 5.53 GeV at 
$T=T_c$ and becomes 11.7 GeV at $T=T_1$, and approaches 
to the zero temperature value 125.7 GeV as the universe 
cools down. Of course, these cosmologically produced 
Higgs particles will quickly decay and disappear.  

Similarly, the W-boson which was massless at high 
temperature becomes massive at $T= T_c$ when 
the Higgs field acquires the non-vanishing vacuum 
expectation value. So, as the Higgs vacuum $\rho=0$ 
starts to tunnel to $\rho=\rho_+$ at $T_c$, 
the W-boson starts to become massive toward the value
$g\rho_+(T_c)/2\simeq$ 7.1 GeV.  And it acquires 
the mass 10.6 GeV at $T=T_1$, which approaches 
the well-known zero-temperature value 80.4 GeV 
at $T=0$. 

\begin{figure}

\includegraphics[height=4.5cm, width=7cm]{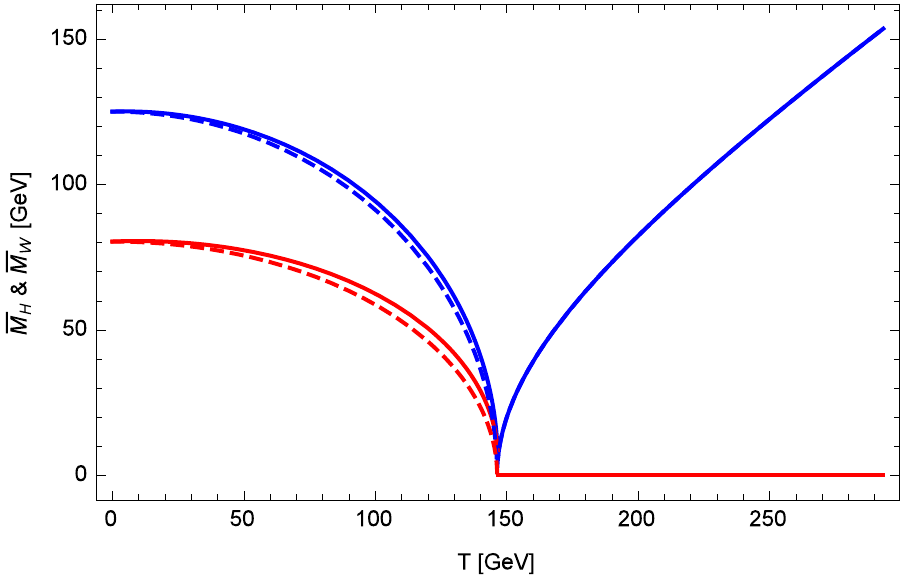}
\caption{\label{hwmass} The temperature-dependent 
Higgs and W-boson masses. The blue and red curves 
represent the Higgs and W-boson masses, and the solid 
and dotted lines represent the masses in the first 
and second order and phase transitions.}
\end{figure}

On the other hand, in the second order phase transition 
we have from (\ref{epot2})
\begin{gather}
\bar M_H^2
=\left\{\begin{array}{ll} 
\big[(T/T_c)^2 -1 \big] M_H^2/2, &~~~T \ge T_c,  \\
\big[1 -(T/T_c)^2 \big] M_H^2,  &~~~T < T_c, 
\end{array} \right.
\end{gather}
so that the Higgs mass becomes zero at $T_c$. This 
makes an important difference in the monopole 
production density, as we will see soon. 
The temperature-dependent Higgs and W-boson masses
are shown in Fig. \ref{hwmass}, where the blue and 
red curves represent the Higgs and W-boson masses,
and the solid and dotted lines represent the first 
and second order and phase transitions. 

\section{Electroweak Monopole Production after Phase Transition}

The production of magnetic monopole by the phase transition 
in early universe was discussed by Kibble and others, and later 
by Zurek who refined the Kibble's estimate of the monopole 
density taking into account the dynamics of equilibrium 
processes \cite{kibble,pres,guth,zurek}. But before we discuss 
the cosmological electroweak monopole production, we briefly 
review the big bang cosmology in early universe. 

The big bang cosmology is based on the cosmological principle
which assumes that the universe is isotropic and homogeneous,
and is described by the Robertson-Walker metric which 
implements this principle
\begin{gather}
ds^2=-dt^2+ R^2(t) \Big(\frac{dr^2}{1-kr^2}  \nn\\
+r^2(d\theta^2+\sin^2 \theta d\varphi^2) \Big),
\label{rwm}
\end{gather}
where $R$ is the scale factor and $k=\pm1,~0$ represents 
the curvature (closed, open, and flat) of the universe. 
Coupled to the perfect fluid energy-momentum tensor 
$T_\mn=(\varrho+p)\xi_\mu \xi_\nu+p g_\mn$, the metric 
gives the familiar Friedmann equation 
\begin{gather}
H^2+\frac{k}{R^2}=\frac{8\pi}{3} G_N \varrho,  \nn\\
\dot \varrho +3 H (\varrho+p)=0,
\label{bbeq}
\end{gather}
where $H=\dot R/R$ is the Hubble parameter and 
$\varrho$, $p$, and $\xi_\mu$ are the density, 
pressure, and the 4-velocity of the perfect fluid. 
In the big bang cosmology one also assumes 
that the expansion is adiabatic,
\begin{gather}
\frac{d}{dt} (sR^3)=0,
\end{gather}
where $s$ is the entropy density of the universe.

The metric (\ref{rwm}) tells that the coordinate distance 
the light (or a massless particle) which leaves $r=0$ 
at $t=0$ and arrives at $r$ after the time $t$ is 
\begin{gather}
d(t)=\int_{t_0=0}^{t_1=t} \frac{dt'}{R(t')}
=\int_{0}^{r} \frac{dr'}{(1-kr'^2)^{1/2}},
\end{gather}
so that the horizon distance of the universe at
$t$ is given by  
\begin{gather}
d_H(t)=R(t)\int_0^{t} \frac{dt'}{R(t')}=R(t)d(t).  
\end{gather}
Moreover, letting the light which leaves $\delta t_0$ later 
and travels the same coordinate distance arrive $\delta t_1$ 
later, we have 
\begin{gather}
\frac{\delta t_0}{R(t_0)}=\frac{\delta t_1}{R(t_1)}.
\end{gather}     
From this we can define the ``redshift" parameter $z$ 
by the ratio of the detected wavelength at $r_1$ to 
the emitted wavelength at $r=0$,
\begin{gather}
1+z= \frac{\lambda_1}{\lambda_0}
=\frac{R(t_1)}{R(t_0)}, 
\end{gather}  
which provides us an important means to test the change 
of the scale factor $R(t)$ experimentally. 

To solve the Friedmann equation we need to specify 
the equation of state of the matter. At high temperature 
the universe is in the radiation dominant era, and we 
may assume that the matter is made of ideal quantum 
gas of massless particles whose energy density 
$\varrho$ and entroty density $s$ are given by 
\begin{gather}
\varrho=3p=\frac{\pi^2}{30} g_*(T) T^4,
~~~~~s=\frac{2\pi^2}{45} g_s(T) T^3,  \nn\\
g_* = \sum_\text{bosons} g_{bi}  \Big(\frac{T_i }{T}\Big)^4
+\frac{7}{8} \sum_\text{fermions} g_{fi} 
 \Big(\frac{T_i }{T}\Big)^4,   \nn\\
g_s = \sum_\text{bosons} g_{bi}  \Big(\frac{T_i }{T}\Big)^3
+\frac{7}{8} \sum_\text{fermions} g_{fi} 
 \Big(\frac{T_i }{T}\Big)^3,
\label{dsrdom}
\end{gather}
where $g_i$ and $T_i$ are the internal degrees and 
the temperature of the $i$-th relativistic particle. 

From this we have (when the temperature is not near 
any mass threshold)
\begin{gather}
R=\Big(\frac{30 E}{\pi^2 g_*} \Big)^{1/4}~\frac{1}{T}
=\Big(\frac{45 S}{2 \pi^2 g_s} \Big)^{1/3}~\frac{1}{T},
\label{RTeq}
\end{gather}
where $E=\varrho R^4$ and $S=s R^3$ are the integration 
constants which represent the total energy and entropy
of the universe. 

With this (\ref{bbeq}) is written as
\begin{gather}
\Big(\frac{\dot T}{T}\Big)^2+\delta(T) T^2
=\frac{4\pi^3}{45} G_N g_* T^4,  \nn\\
\delta(T)=\frac{k}{R^2T^2}
=k~\Big(\frac{\pi^2 g_*}{30 E}\Big)^{1/2}  \nn\\
=k~\Big(\frac{2\pi^2 g_s}{45 S}\Big)^{2/3}.
\label{bbeq2}
\end{gather}
On the other hand, in the radiation dominant era, 
the curvature term $\delta(T)$ become negligible 
compared to the density term when $R$ becomes small, 
so that we may assume $k=0$. With this we can solve 
(\ref{bbeq2}) and find 
\begin{gather}
T^2\simeq \frac{C m_p}{2t},
\label{T}
\end{gather}
where $C=\sqrt{45/4\pi^3 g_*(T)}$ and 
$m_p\simeq 1.2 \times 10^{19}$ GeV is the Planck mass. 
This, with (\ref{RTeq}) gives  
\begin{gather}
R(t)=\Big(\frac{30 E}{\pi^2 g_*} \Big)^{1/4} 
\sqrt{\frac{2t}{C m_p}},  \nn\\
H =-\frac{\dot T}{T} \simeq \frac{T^2}{Cm_p} 
\simeq \frac1{2t},
\label{R}
\end{gather}
This remains a good approximation for $T_e<T$, where
$T_e \simeq 0.26~{\rm eV}$ is the temperature where 
the matter and radiation become equal.

From (\ref{bbeq}) we have 
\begin{gather}
\Omega=\frac{\varrho}{\varrho_c}
=1+\frac{k}{H^2R^2},  
~~~\varrho_c=\frac{3H^2}{8\pi G_N}.
\end{gather}
Notice that $\Omega$ (as well as $\varrho$ and $\varrho_c$)
is time-dependent. Since $H^2=(8\pi G_N/3) \varrho_c$, 
this can be written as
\begin{gather}
\Omega-1=\frac{k}{H^2 R^2} 
\propto \frac{1}{T^2}.
\end{gather}
This means that $k/H^2R^2$ must have been extremely small 
in the early universe. But this is very strange because 
this curvature term determines the fate of the universe
in the later stage. This, of course, is the flatness problem. 

From (\ref{R}) we have the horizon distance given by
\begin{gather}
d_H(t)=R(t)\int_0^t \frac{dt'}{R(t')}=2t
=\frac{Cm_p}{T^2}.
\end{gather} 
At the electroweak temperature all particles of 
the standard model contribute to $g_*$, so that 
we have $g_* \simeq 106.75$ and $C\simeq 0.058$. 
This tells that at $T=T_c$ the universe was roughly 
in $1.1 \times 10^{-11} \sec$ after the big bang, 
and the horizon distance was about $d_H\simeq 1/H 
\simeq Cm_p/T_c^2 \simeq 0.65~{\rm cm}$. 

The cosmological principle presupposes that the universe 
is homogeneous. But obviously this does not mix well with 
the causality. To see this notice that the horizon size and 
the size of the universe fixed by the scale factor $R$ need 
not be the same. From (\ref{T}) and (\ref{R}) we have 
\begin{gather}
\Big(\frac{d_H(T)}{R(T)}\Big)^3
=\Big(\frac{\pi^2 g_*}{30E } \Big)^{3/4} 
\Big(\frac{Cm_p}{T}\Big)^3 \propto \frac1{T^3}.
\end{gather}
So, normalizing $R$ by $d_H(T_e)=R(T_e)$, we can 
express the ratio of the horizon volume $V_H$ to 
the actual volume $V$ of the universe by (when $T_e<T$)
\begin{gather}
\Big(\frac{V_H(T)}{V(T)}\Big)^3
=\Big(\frac{T_e}{T} \Big)^3.
\end{gather}
So most of the universe which was visible and homogeneous 
at the matter-radiation eqality temperature were causally 
disconnected in the early universe. But it is difficult to 
understand how the causally disconnected universe became 
homogeneous at $T_e$. This is the essence of the horizon 
problem in the big bang cosmology. 

Now we are ready to discuss the electroweak monopole 
production, or more precisely the monopole-antimonopole 
pair production, since they have to be crated in pairs. 
There have been many works on the monopole productions 
in the literature, but most of the discussions were on 
the grand unification monopole \cite{kibble,pres,guth,zurek}. 

In general the monopole production mechanism depends 
on the type of the phase transition. There are two key 
factors which determine the initial monopole density, 
the time of the monopole formation and the correlation 
length of the phase transition at this time. And they 
depend on the type of phase transition. 

In the second order phase transition the monopole 
formation is assumed to take place around the critical 
temperature. But in the first order phase transition 
the monopole formation is assumed to take place below 
the critical temperature, during the quantum tunneling 
through the vacuum bubble collisions. So the monopole 
production mechanism in two cases is totally different. 

As we have pointed out, however, the electroweak 
phase transition is mildly the first order. This 
makes the electroweak monopole production more 
complicated \cite{and,koba}. So it is worth to 
discuss the electroweak monopole production in both 
the first order and second order phase transitions, 
and we discuss the two cases separately. 

\subsection{Monopole production in the second order 
phase transition}

In the phase transition the average distance between 
two topological defects is given by the correlation 
length, so that the initial monopole density is 
determined by the correlation length $\xi$ which is 
set by the Higgs mass, $\xi=1/\bar M_H$. But in 
the second order phase transition $\xi$ becomes 
infinite since the Higgs mass becomes zero at $T_c$. 
In this case the only parameter which sets the length 
scale at $T_c$ and can play the role of the correlation 
length is the horizon distance, and from the causality 
argument Kibble has proposed that the initial density 
of the monopole (and anti-monopole) must be bounded by 
the horizon distance \cite{kibble}
\begin{gather}
\Big(\frac{n_m}{T^3} \Big)_i  \gtrsim 
\Big(\frac{T_c}{  C m_p} \Big)^3 
\simeq 8.7 \times 10^{-48}.
\label{Kbound}
\end{gather}
This Kibble bound has provided an important limit on 
the monopole density in the second order phase 
transition. 

The phase transition, however, does not take place 
instantaneously but continuously, so that we have to 
incorporate the dynamics of equilibrium process.
And the Kibble bound was refined by Zurek who took 
care of the relaxation time $\tau$ in the phase 
transition,
\begin{gather}
\tau = \frac{\tau_0}{|\varepsilon(T)|^{\mu}}, 
~~~\xi = \frac{\xi_0}{|\varepsilon(T)|^{\nu}},  \nn\\
\varepsilon(T) = (T_c - T)/T_c,
\end{gather}
where $\varepsilon(T)$ is related with the quenching 
time scale $\tau_Q = t/\varepsilon$ which is related 
to $ 1/ H(T_c)$ in the cosmological context. Also 
the critical exponents $\mu$ and $\nu$ characterize 
the universality class of the transition, and $\tau_0$ 
and $\xi_0$ are dimensional parameters determined 
by the microphysics. In our case they are given by 
$\xi_0 \approx \tau_0 \sim 1/(\sqrt{\lambda}T_c)$.

As the temperature approaches the critical value, 
$\tau$ becomes sufficiently longer and the process 
critically slows down. However, $\xi$ increase 
indefinitely but the propagation of small fluctuations 
is finite (limited by the speed of light). Therefore, there 
is a characteristic  time $t_*$ when the correlation 
length freezes. From these the correlation length is 
given by $\xi(t_*) =\xi_0 |\tau_0/\tau_Q|^{\nu/(1+\mu)}$. 
The causality $\xi \le c\tau$ implies $\nu \le \mu$ 
and we will assume $\mu=\nu$ from now on. 
 
With $\tau_Q \simeq  H^{-1}(T_c) = Cm_p / T_c^2 $  
the relic density of monopole following from the Zurek 
mechanism in the second order phase transition is 
expressed by
\begin{gather}
\Big(\frac{n_m}{T^3}\Big)_i
\simeq\frac{ g_P}{\xi^3 (t_*) T_c^3}  \nn \\
\simeq 0.02 \times \Big(\frac{28.4 T_c}{m_p}\Big)^p,
~~~~p={\frac{3\nu}{1+\nu}},
\label{sorder}
\end{gather}
where $g_P$ is a geometrical factor of oder one-tenth.
From this we can estimate the initial monopole production 
density 
\begin{gather}
8.4 \times 10^{-22} \lesssim \Big(\frac{n_m}{T^3} \Big)_i
\lesssim 3.6\times 10^{-18},
\label{Zbound}
\end{gather}
where we have used critical exponent $\nu$ which is 
allowable in the field theory model, $0.5\le \nu \le 0.7$. 
The main difference between the Zurek result and 
the Kibble's bound is the power $p$ of the $(T_c/m_p)^p$ 
suppression factor. So, when $p$ is small, monopoles 
with relatively light mass can contribute to the dark 
matter substantially.

Actually there is a more realistic way to estimate 
the initial monopole density. To figure out the monopole 
density more accurately, the important thing to know 
is the time of the monopole formation. To find this, 
notice that the creation of the monopole requires 
the change of topology, in particular the appearance 
of zero points of the Higgs vacuum which become 
the seeds of the monopoles. Clearly these zero points 
do not appear instantaneously at the critical temperature, 
but are naturally induced by the thermal fluctuations 
of the Higgs field after the phase transition. This 
means that the monopoles do not appear at the critical 
temperature, but some time later.  

To find the monopole production temperature, notice 
that, just below the critical temperature the Higgs 
field is still subject to large fluctuations which 
bring $\langle \rho \rangle$ back to zero.  This is 
possible so long as
\begin{gather}
\xi^3 \Delta F \le  T,
\label{flcon}
\end{gather}
where $\xi(T)$ is the correlation length and 
$\Delta F(T)=V(\rho_s)-V(\rho_{+})$ is the difference 
in free energy density between two phases with
different symmetry. This is because the fluctuation 
energy should not exceed the thermal energy. 

The temperature at which the equality holds in 
(\ref{flcon}) is the Ginzberg temperature $T_G$. 
In the second order approximation (\ref{epot2})
we have $\bar{M}_H^2=[1-(T/T_c)^2]M_H^2 $, 
so that the Ginzburg temperature is given by
\begin{gather}
T_G = \frac{T_c}{\sqrt{1+ 32\lambda^2/C_2}}
\simeq 56.0~{\rm GeV},  
\label{Gtemp2}
\end{gather} 
which is lower than the critical temperature $T_c$.
The correlation length $\xi_G$ which saturates the equality 
in (\ref{flcon}) at the Ginzburg temperature is given by
\begin{gather}
\xi_G 
=\frac{\sqrt{1+ 32 \lambda^2/C_2} }{8\lambda T_c}
\simeq 1.7 \times 10^{-16}~{\rm cm}.
\label{col2}
\end{gather}
This is well within the horizon distance. 

Now, assuming that the monopoles are produced between 
$T_c$ and $T_G$ we may choose the correlation length 
$\xi_i$ which determines the initial monopole density 
and the corresponding monopole production temperature 
$T_i$ to be 
\begin{gather}
\xi_i=\xi(T_i) \simeq 2.2 \times 10^{-16}~\text{cm},  \nn\\
~~~~T_i=\frac{T_G+T_c}{2} \simeq 101.2~\text{GeV}.
\end{gather}
Now, let $g_P$ be the probability that one monopole 
is created inside a domain of size $\xi_i$. With this 
we find a new initial monopole density which can 
replace the Zurek bound,  
\begin{gather}
\Big(\frac{n_m}{T^3}\Big)_i 
\simeq \frac{g_P}{\xi_i^3 T_i^3}
\simeq 7.1\times 10^{-2},
\label{C2bound}
\end{gather}
where we have assumed $g_P\simeq 0.1$. This is bigger 
than the Zurek bound (\ref{Zbound}) roughly by the factor 
$10^{16}$, which shows that the initial monopole density 
depends crucially on how to estimate it. 

\subsection{Monopole production in the first order 
phase transition}

The above pictures, however, may not properly describe 
the electroweak monopole production. This is because 
the Kibble and Zurek argument does not apply since
the electroweak phase transition is the first order. When 
the phase transition is of the first order, the correlation 
length becomes finite because the Higgs mass becomes 
non-zero at the critical temperature. Moreover, the potential 
barrier between the symmetric and broken vacua modifies 
the Ginzburg temperature. 

In the strongly first order phase transition, 
the symmetric vacuum becomes meta-stable
below the critical temperature, and bubbles of 
broken phase will nucleate and expand. Within 
each bubble the Higgs field is correlated, but 
there is no correlation in Higgs field in difference 
bubbles. Thus when the bubbles collide, they can 
create the monopoles whose density is proportional to 
the density of bubbles. Since the expansion speed 
cannot exceed the speed of light, the size of 
a bubble is also limited by the horizon distance. 
So we have the following bound on the relic 
monopole abundance \cite{guth} 
\begin{gather}
\Big(\frac{n_m}{T^3}\Big)_i \gtrsim
\Big[\frac{T_c}{C m_p} \log \Big(\frac{Cm_p}{T_c}\Big)^4\Big]^3  \nn\\
\simeq 2.6 \times 10^{-41}.
\label{Wbound}
\end{gather}
Notice that the logarithmic factor gives the enhanced 
monopole glut for the first order phase transition.

However, the electroweak phase transition is very mildly 
first order, so that the bubble formation and thus the monopole 
production by the bubble collisions becomes unimportant 
and negligible. Indeed, in the first order phase transition 
described by (\ref{epot}), we may neglect the tunneling of 
the potential barrier and assume that the monopole formation 
takes place after the phase transition is completed, between 
$T_1$ and the Ginzburg temperature. So we have to find
the Ginzburg temperature first.
 
\begin{figure}
\includegraphics[height=4.5cm, width=7cm]{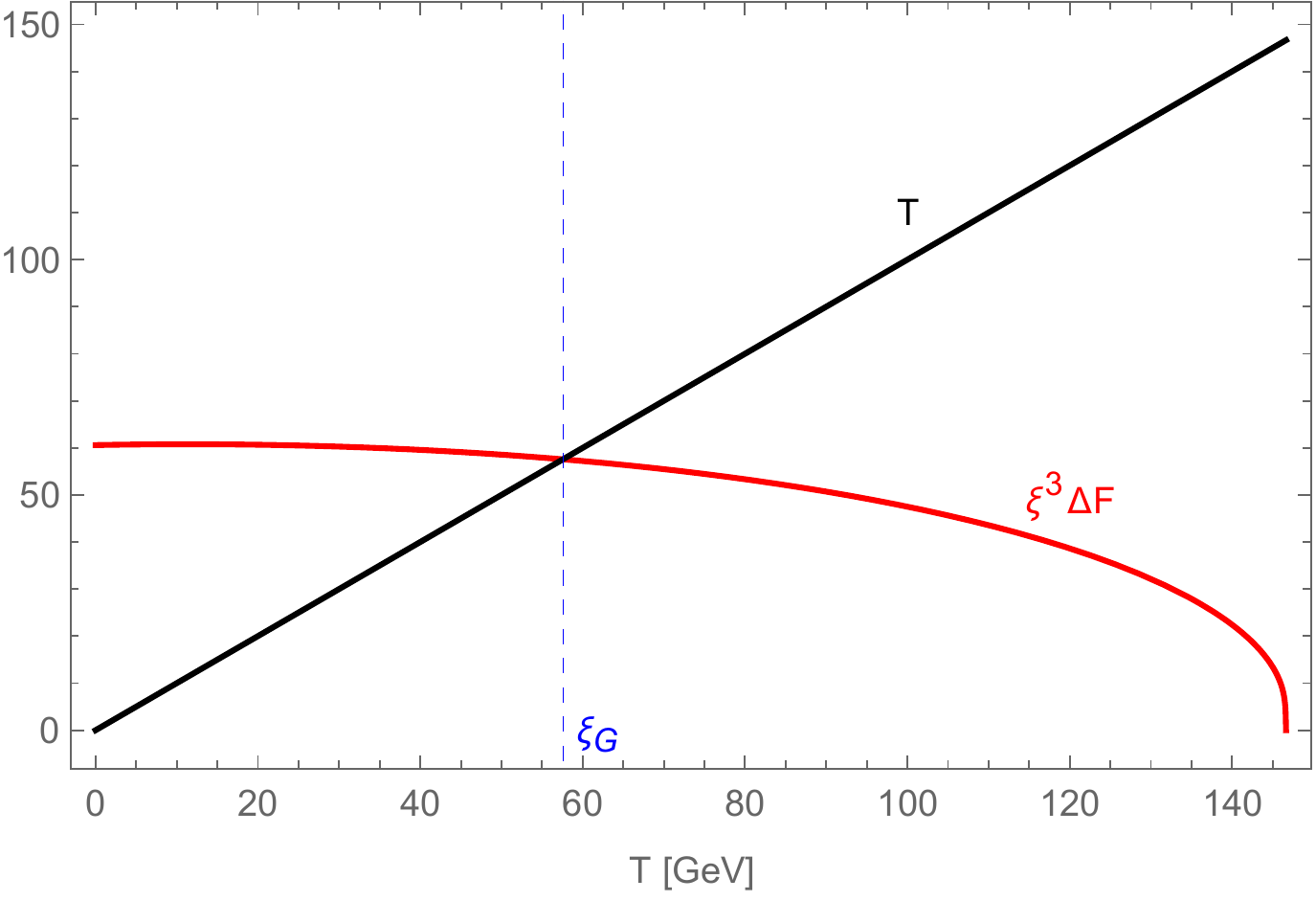}
\caption{\label{Gtemp} The determination of the Ginzburg 
temperature in the first order electroweak phase transition. 
Here the red and blue curve represents $\xi^3 \Delta F$ 
and the black line represents $T$.}
\end{figure}

Unfortunately in this case the Ginzburg temperature 
can not be expressed in a simple form. But we can 
calculate $T_G$ numerically directly from the effective 
potential (\ref{epot}). With $\bar{M}_H^2=\lambda (\rho_+^2 
+\rho_0^2 )/2-C_2 T^2$ we plot $\xi^3 \Delta F$ 
and $T$ in red curve in Fig. \ref{Gtemp}. From this we find
\begin{gather}
T_G \simeq  57.6~\text{GeV},  \nn\\
\xi(T_G) \simeq 1.7\times 10^{-16}~\text{cm},
\label{Gtemp1}
\end{gather} 
which is almost identical to (\ref{Gtemp2}) and (\ref{col2}). 
This tells that, on the Ginzburg temperature and 
the correlation length, the second order approximation 
works quite well.

With this we may assume that the period of the monopole 
formation is between $T_1$ and $T_G$, 
\begin{gather}
\bar t=t(T_G)-t(T_1)=\frac{Cm_p}{2T_G^2}
-\frac{Cm_p}{2T_1^2}  \nn\\
=\frac{Cm_p}{2} \times \frac{T_1^2-T_G^2}{(T_1 T_G)^2}
\simeq 6.0\times 10^{-11}~{\rm sec}.
\end{gather} 
And during this period the Higgs vacuum must fluctuate 
to zero to create the monopoles. 

We can estimate the time scale of the fluctuation $\Delta t$ 
from the uncertainty principle $\Delta E\cdot \Delta t 
\simeq 1$, where $\Delta E$ is given by the Higgs vacuum 
$\rho_+$ around this time,
\begin{gather}
\Delta t \simeq \frac1{\Delta E}  
\simeq 5.0\times 10^{-27}~{\rm sec}, \nn\\
\Delta E \simeq \frac{\rho_+(T_1)+\rho_+(T_G)}{2}.
\end{gather}
From this we can estimate the number of the fluctuations 
of the Higgs field,
\begin{gather}
N\simeq \frac{\bar t}{\Delta t}\simeq Cm_p \times 
\frac{(T_1^2-T_G^2)(\rho_+(T_1)
+\rho_+(T_G))}{(T_1 T_G)^2} \nn\\
\simeq 1.2\times 10^{16}.
\end{gather}
This tells that there is ample time for the vacuum fluctuation
of the Higgs field to produce the monopoles. From this we
conclude that the dominant mechanism for the electroweak 
monopole production is not the bubble collisions during 
the phase transition but the quantum fluctuation of 
the Higgs vacuum after the phase transition.     

From (\ref{Gtemp1}) we can estimate the initial density 
of the monopoles in the first order phase transition.
Assuming that the monopoles are produced between 
$T_1$ and $T_G$, we have 
\begin{gather}
\Big(\frac{n_m}{T^3}\Big)_i
\simeq \frac{g_P}{\xi_i^3  T_i^3}
\simeq 9.1\times 10^{-4},  \nn\\
\xi_i=\frac{\xi(T_1)+\xi(T_G)}{2}
\simeq 9.3 \times 10^{-16}~\text{cm},  \nn \\
T_i=\frac{T_1+T_G}{2}\simeq 102.0~\text{GeV}.
\label{C1bound}
\end{gather} 
Again this is a huge number compared with (\ref{Zbound}) 
or (\ref{Wbound}). Moreover, this is smaller than 
(\ref{C2bound}) by the factor $10^{-2}$, which tells 
that the second order approximation is not so trustable 
to estimate the monopole density. This is the main 
difference between our estimate and the old estimates. 

Obviously, the monopole production consumes energy, 
so that we must know how much energy it consumes.
If it consumes too much energy, it could cause us trouble. 
In fact, it is well known that in the grand unification 
the superheavy monopoles are produced too copiously 
that they force the universe supercool. Moreover, 
the mass of the remnant monopoles dominates 
the mass of all other matters by many orders of 
magnitude \cite{pres, guth}. Clearly, this is incompatible 
with the standard cosmology. So we have to check if 
the electroweak monopoles could cause a similar trouble.  
       
To see how much energy we need to produce 
the electroweak monopoles, we can calculate 
the energy density of the monopoles from 
(\ref{C1bound})
\begin{gather}
\varrho_{mo} (T_i)=M_m  (n_m)_i  
\simeq 9.0\times 10^{-3}  ~T_i^4
~\Big(\frac{M_m}{1~\text{TeV}}\Big),
\label{meden}
\end{gather}
where the mass of electroweak monopole is assumed 
to be around 1 TeV. This is because the monopole 
mass must be about 100 times bigger than the W-boson 
mass \cite{epjc15,ellis,plb16}. This should be compared 
with the total energy density of the universe (\ref{dsrdom}) 
at $T_i$
\begin{gather}
\frac{\varrho_{mo} (T_i)}{\varrho(T_i)}
\simeq \frac{0.268}{\pi^2 g_*(T_i)} 
\Big(\frac{M_m}{1~\text{TeV}}\Big) \nn\\
\simeq 2.5\times 10^{-4}\Big(\frac{M_m}{1~\text{TeV}}\Big).
\label{medenratio}
\end{gather}
This tells that the universe need to consume only 
a tiny fraction (about $0.025~\%$) of the total 
energy to produce the monopoles.  This is nice, 
because this assures that the monopole production 
does not alter the standard cosmology. 

This is very important because, had the monopoles 
been produced too copiously, their energy density 
would have dominated the total energy density of 
the universe. In fact, this has been a major problem 
with the grand unification monopole \cite{kibble,pres}. 
The above result assures that the electroweak monopole 
has no such problem. 

To see the importance of this observation, suppose 
$\xi_i$ in (\ref{C1bound}) were 10 times bigger. 
In this case $n_m$ would become 1000 times bigger, 
so that the universe would have to consume about 
a quarter of the total energy to produce the monopoles. 
This would have been a serious problem, because 
this might supercool the universe and could alter 
the standard cosmology. 

In fact, we can say that even (\ref{C1bound}) is 
an overestimation. The reason is that, as we will 
see soon, the monopole-antimonopole capture 
radius becomes much bigger the correlation length 
$\xi_i$, so that most of the monopoles annihilate 
with the anti-monopoles as soon as they are produced.
Moreover, this ahnnihilation continues very long 
time, till the universe cools down to about 29.5 MeV. 
This is basically because the monopoles couple 
strongly, i.e., magneticically.  

Fortunately we do not have to worry about this 
overestimate of the initial monopole density. This 
is because, as we will see, the final density of 
the monopoles turns out to be rather insensitive 
to the initial monopole density.  

\section{Density of Relic Electroweak Monopoles}

Since the magnetic charge of the monopoles are topologically 
conserved, they are absolutely stable. So, once created, 
the monopoles do not decay. As we have pointed out, 
however, the initial monopole density changes. There 
are two factors which make this change, the Hubble 
expansion and the annihilation of monopole-antimonopole 
pairs, and the evolution of monopole (and anti-monopole) 
density $n_m$ is determined by the Boltzmann 
equation \cite{pres,zel}
\begin{gather}
\frac{d n_m}{dt} + 3 H n_m = -\sigma n_m^2.
\label{boltz1}
\end{gather}
where $H$ and $\sigma$ are the Hubble parameter and
the monopole-antimonopole annihilation cross section. 

Obviously the Hubble expansion dilutes the monopole 
density, but the monopole-antimonopole annihilation 
also deflates the monopole density. And this annihilation 
process becomes very important for us to estimate 
the remnant monopole density at present universe. 

To study this annihilation process, we have to find out 
the monopole-antimonopole annihilation cross section 
first. The annihilation is controlled by two competing
forces, the thermal Brownian motion (random walk) in 
hot plasma of charged particles and the attraction 
between the monopoles and anti-monopoles. After 
the creation the monopoles diffuse in a hot plasma 
of relativistic charged particles by the Brownian motion 
with the mean free path $l_{\rm free}$,
\begin{gather}
l_{\rm free} =v_T t_{\rm free} \simeq {\sqrt \frac{ T}{M_m}} 
\times \frac{M_m}{T\sum_i n_i \sigma_i} \nn\\
=\frac{1}{BT} \sqrt{\frac{ M_m}{T}},  
~~~~~B =\frac{1}{T}{\sum_i n_i \sigma_i},
\label{lfree}
\end{gather}
where $v_T\simeq \sqrt{T/M_m}$ and $t_\text{free}$ 
are the thermal velocity and the mean free time of 
the monopoles, $n_i$ and $\sigma_i$ are the number 
density and the cross section of the $i$-th relativistic 
charged particles and the sum is the sum over all spin 
states \cite{pres}. 

Notice that here we have assumed that the Brownian 
motion of the monopoles is non-relativistic. This is 
justified by the fact that the monopoles produced (with 
the initial mass around $1~\text{TeV}$) can be treated 
as non-relativistic particles, even though the universe 
is still in the radiation dominant era.  

With
\begin{gather}
n_i \simeq \frac{3\zeta(3)}{4\pi^2} T^3,
~~~\sigma_i \simeq \Big(\frac{g_m q_i}{4\pi}
\Big)^2 \frac{1}{T^2}
= \Big(\frac{q_i}{e}\Big)^2 \frac{1}{T^2},
\end{gather}
we have 
\begin{gather}
B\simeq \frac{3\zeta(3)}{4\pi^2} \sum_i 
\Big(\frac{q_i}{e}\Big)^2 \simeq 0.09 
\times \sum_i \Big(\frac{q_i}{e}\Big)^2,
\end{gather}
where $q_i$ is the electric charge of the $i$-th particle
and $\zeta(3)=1.202...$ is the Riemann zeta function. 
Since the charged particles in the plasma are the leptons 
and quarks, we may put $B\simeq 3$. 

With this we can estimate the thermal velocity and 
the mean free length of the monopoles around 
the temperature $T_i$. With (\ref{C1bound}) we have
$v_T\simeq 0.31 \times (M_m/1~\text{TeV})^{-1/2}~c$ and  
$l_{\rm free} \simeq 2.0 \times (M_m/1~\text{TeV})^{1/2}
\times 10^{-16}~{\rm cm}$ at $T_i$. 

Now, against the thermal random walk of the monopoles, 
we have the attractive Coulomb force between monopoles 
and anti-monopoles which makes them drift towards each 
other. The drift velocity $v_d$ of the monopole at a distance 
$r$ from the anti-monopole is given by \cite{vil}
\begin{gather}
v_d \simeq \frac1{\alpha} \times \frac1{B T^2  r^2}
= \frac{\alpha_m}{B T^2  r^2},
\end{gather}
where $\alpha=e^2/4\pi$ is the electromagnetic fine 
structure constant and $\alpha_m$ is the monopole 
fine structure constant defined by $\alpha_m=g_m^2/4\pi$,
where $g_m=4\pi/e$ is the magnetic charge of the monopole.
Notice that $\alpha_m=1/\alpha$. 

Clearly the drag force generated by the Coulomb attraction 
can dissipate enough energy for the monopoles to be 
captured by the nearby anti-monopoles. So we have 
the monopole-antimonopole bound states which quickly 
annihilate, when the mean free path becomes less than 
the capture radius
\begin{gather}
l_{\rm free} \le r_\text{capt} =\frac{\alpha_m}{T},
\end{gather}
where $r_\text{capt}$ is determined by the condition that 
the thermal energy is equal to the potential energy of 
the monopole-antimonopole pair. With (\ref{C1bound}) we 
have $r_\text{capt}\simeq 2.7 \times 10^{-14}~\text{cm}$ 
at $T_i$. Notice that the capture radius is much bigger 
(about hundred times) than the mean free length and 
correlation length, which indicates that the monopoles 
are annihilated as soon as they are produced. This is 
a clear evidence that (\ref{C1bound}) could be 
an overestimation.

There is another strong evidence to support this 
observation. Notice that the drift velocity becomes 
$v_d  (r=\xi_i)\simeq 2~c$. This, of course, is 
an  unrealistically large and impossible number. 
But this does tell that the Coulomb attraction 
between the monopoles and anti-monopoles is 
much more stronger than the thermal diffusion 
around $T_i$. This assures that the annihilation 
is much more important than the diffusion around 
$T_i$, so that as soon as they are created, they 
are annihilated. This shows that the initial monopole 
density (\ref{C1bound}) is indeed an overestimation.   

\begin{figure}
\includegraphics[height=4.5cm, width=7cm]{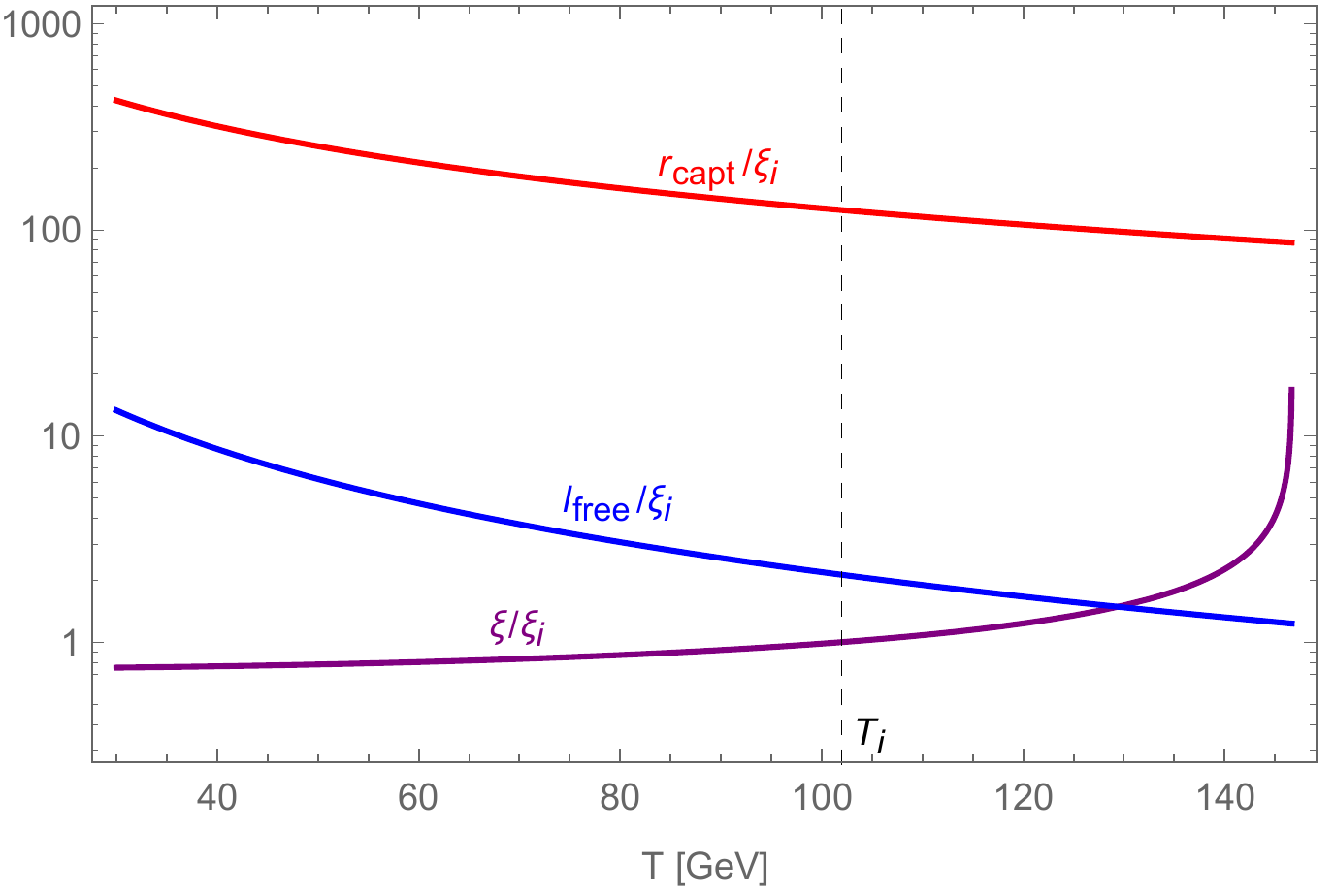}
\caption{\label{scales} The relevant scales, $\xi$ in 
purple, $l_{\rm free}$ in blue, and $r_\text{capt}$ 
in red, against $T$. They are normalized by 
the correlation length $\xi_i$ at $T_i$. Here we set 
$M_m=5~{\rm TeV}$.}
\end{figure}

In Fig. \ref{scales}  we plot the relevant scales $\xi$, 
$l_{\rm free}$, and $r_\text{capt}$ against $T$ for 
comparison. This confirms that the capture radius is 
much bigger than the mean free length and correlation 
length in a wide range of $T$, and reassures that 
the monopole-antimonopole annihilation is much 
more stronger than the monopole production.  

Now, if we let the mean distance between 
the monopole and anti-monopole $r$ be 
$r\simeq n_m^{-1/3}$, the capture time is 
given by
\begin{gather}
t_\text{capt}\simeq \frac{r}{v_d} 
\simeq \alpha \times \frac{B T^2}{n_m}.
\end{gather}
From this we have the monopole-antimonopole 
annihilation cross section 
\begin{gather}
\sigma \simeq \frac{1}{t_\text{capt} n_m} 
\simeq \frac{\alpha_m}{B T^2}.
\end{gather}
Notice that, with (\ref{C1bound}) we have 
$t_\text{capt} \simeq 1.6 \times 10^{-19}~\text{sec}$ 
and $\sigma \simeq 1.7 \times 10^{-30}~\text{cm}^2$
at $T_i$. 

With this we can solve the Boltzmann equation 
(\ref{boltz1}). In term of $\tau=M_m/T$ the Boltzmann 
equation becomes
\begin{gather}
\frac{d}{d\tau} \Big(\frac{n_m}{T^3} \Big) 
=-\frac{\sigma T^3}{\tau H}\Big(\frac{n_m}{T^3}\Big)^2 \nn\\
=-\alpha_m \times \frac{Cm_p}{BM_m} \Big(\frac{n_m}{T^3}\Big)^2.
\label{boltz2}
\end{gather}
The analytic solution of the Boltzmann equation is 
well-known,
\begin{gather}
\frac{n_m}{T^3} 
 =\frac{1}{A(\tau - \tau_i ) +B}, \nn\\
A=\alpha_m  \times \frac{Cm_p}{BM_m}, 
~~~B=\Big(\frac{n_m} {T^3}\Big)^{-1}_i.
\end{gather}
Notice that when $\tau_i \ll \tau$, only the first term in 
the denominator becomes important. So the monopole 
density approaches to 
\begin{gather}
\frac{n_m}{T^3}\rightarrow \alpha
\times \frac{BT}{C m_p},
\end{gather}
regardless of the initial condition \cite{pres}. 

\begin{figure}
\includegraphics[height=4.5cm, width=7cm]{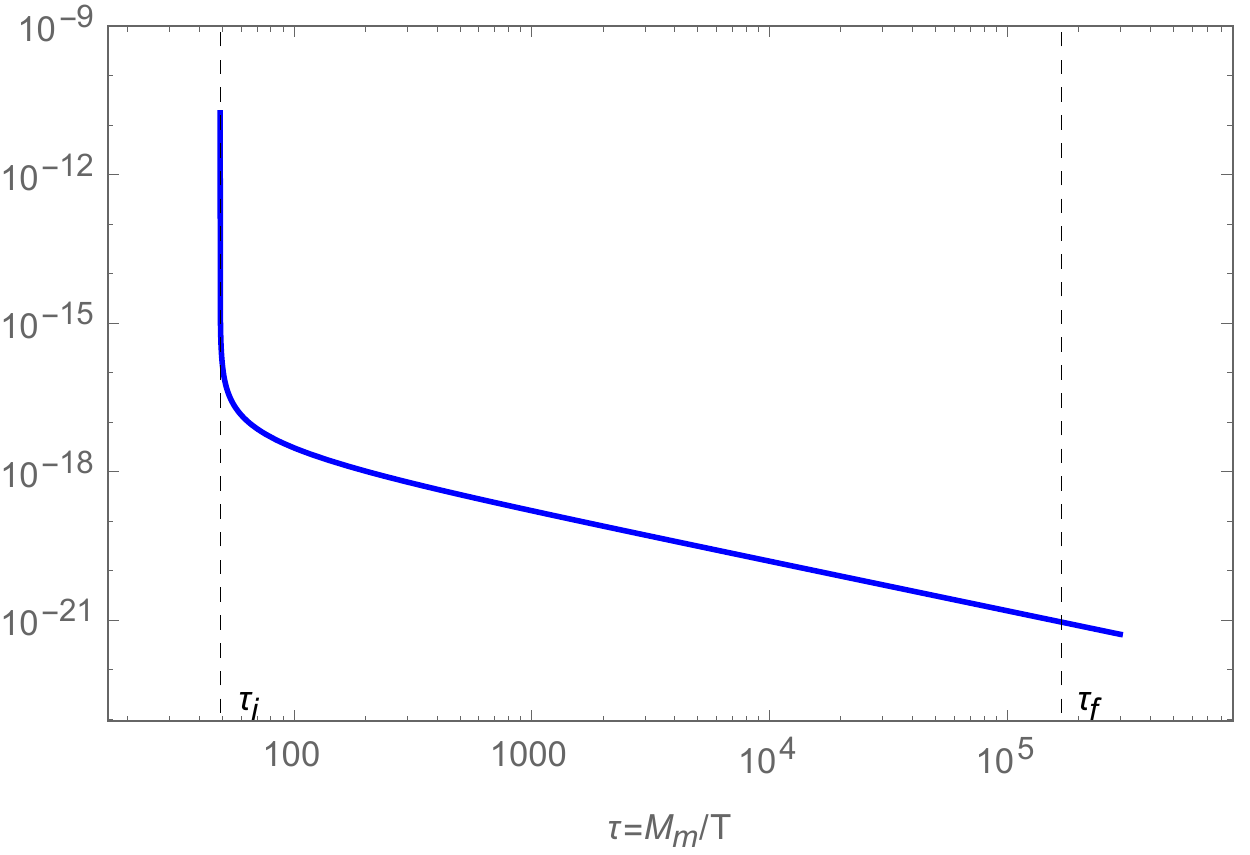}
\caption{\label{mden} The evolution of the monopole 
density $n_m/T^3$ against $\tau=M_m/T$, which shows 
that most of the initial monopoles are quickly annihilated.}
\end{figure}

The evolution of the monopole density $n_m/T^3$ against 
$M_m /T$ is shown in Fig. \ref{mden}, where we have put 
$M_m=5~{\rm TeV}$. This shows that most of the monopoles 
are annihilated as soon as created, after which the annihilation 
continues at a constant rate. Notice that here $B$ and 
the monopole mass $M_m$ are treated as constants, but 
strictly speaking they depends on time. So they should be 
understood as the mean values. 

The diffusive capture process is effective only 
when $\ell_\text{free} < r_\text{capt}$, which 
determines the temperature $T_f$ below which 
the monopole-antimonopole annihilation ceases,
\begin{gather}
T_f  \simeq \alpha^2 \times \frac{M_m}{B^2}
\simeq 5.9 \times  \Big(\frac{M_m}{1~\text{TeV}}\Big)
~{\text {MeV}},
\end{gather} 
so that 
\begin{gather}
\tau_f  =\frac{M_m}{T_f} \simeq (\alpha_m B)^2 
\simeq 1.7 \times 10^5.
\end{gather}
With $M_m=5~{\rm TeV}$, we have $T_f\simeq 29.5~{\rm MeV}$, 
which is below the muon pair annihilation temperature. Actually, 
around this temperature $B$ becomes very small so that 
the capture radius becomes smaller. This is because 
the only charged particles remaining in the plasma are 
electrons and positrons. But the above analysis clearly 
shows that the annihilation continues very long time. 
The reason is basically the monopole-antimonopole 
attraction is much stronger than the electron-positron 
attraction. 

Since $\tau_i \ll \tau_f$ the final density of the monopole 
at $T_f$ is independent of the initial density, and becomes 
\begin{gather}
\Big(\frac{n_m}{T^3}\Big)_f \simeq 
\alpha^3 \times \frac{M_m}{BCm_p} \nn\\
\simeq 1.8\times 10^{-22} 
~\Big(\frac{M_m}{1~\text{TeV}}\Big),
\label{fmden}
\end{gather}
where we have put $B \simeq 3$. Obviously this value is 
much lower than the initial density given by (\ref{C1bound}). 
This confirms that most of the monopoles produced initially 
are annihilated and diluted away. 

As the temperature of universe cools down further, 
the annihilation process becomes unimportant, and 
the number of monopole within the comoving volume 
is conserved thereafter. Notice, however, that around 
$T_f$ the monopoles are still interacting with the electron 
pairs in the hot plasma. But eventually they decouple 
around $T_d\simeq 0.5~\text{MeV}$, when the electron 
pairs disappear and the monopole interaction rate 
becomes less than the Hubble expansion rate.

Assuming that the expansion is adiabatic, we have
\begin{gather}
\frac{n_m}{T^3}=1.8\times 10^{-22} 
\Big(\frac{M_m}{1~\text{TeV}}\Big)
\Big(\frac{g_s(T)}{g_s(T_f)}\Big),
\label{mnden}
\end{gather}
so that the current number and energy density of 
monopole is given by
\begin{gather}
\Big(\frac{n_m}{T^3}\Big)_0
= \Big(\frac{g_{s,0}}{g_{s,f}}\Big)
\Big(\frac{n_m}{T^3} \Big)_{f},  \nn\\
\rho_{mo,0} =M_m  n_{m,0}
= M_m  T_0^3 \Big(\frac{g_{s,0}}{g_{s,f}} \Big)
\Big(\frac{n_m}{T^3} \Big)_{f},
\label{mden0}
\end{gather}
where $T_0 = 2.73~\text{K}=2.35\times 10^{-13}
~\text{GeV}$ is the temperature of the universe today
and $g_{s,f}$ is the effective number of degrees of 
freedom in entropy at $T_f$ shown in (\ref{dsrdom}), 
which is equal to $g_*$ when all the relativistic 
species are in thermal equilibrium at the same 
temperature.

The current density parameter of monopole can be 
written 
\begin{gather}
\Omega_{mo}~h^2 =\frac{\rho_{mo,0}~h^2}{\rho_{c,0}} \nn\\ 
= 1.7 \times 10^{11} \Big(\frac{3.94}{106.75}\Big)
\Big(\frac{n_m}{T^3}\Big)_{f}
\Big(\frac{M_m}{1\text{TeV}} \Big),
\label{dparam}
\end{gather}
where $\rho_{c,0} = {3H_0^2}/{8\pi G_N}$ is the current 
critical density of universe and $h\simeq 0.678$ is 
the scaled Hubble constant defined by 
$H_0/(100~\text{km}~\text{s}^{-1}~\text{Mpc}^{-1})$.
From this and (\ref{mnden}) we have
\begin{gather}
\Omega_{mo}~h^2 \simeq 1.2 \times 10^{-12} 
\times \Big(\frac{M_m}{1~\text{TeV}}\Big)^2.
\label{dparam0}
\end{gather}
With $h\simeq 0.678$ and $M_m\simeq$ 5 TeV, we have
$\Omega_{mo} \simeq 6.53 \times 10^{-11}$. This is about 
$1.31 \times 10^{-9}$ of the baryon density, much less 
than the density of $^3{\rm He}$. This assures that 
the electroweak monopole cannot be a dark matter 
candidate. 

In terms of the number density, this translates to about 
$6.1 \times 10^{-20} ~\text{cm}^{-3}$, or about 
$2.3 \times 10^{-13}~n_b$, where $n_b \simeq 2.5 
\times 10^{-7}~\text{cm}^{-3}$ is the number density 
of the baryons. Intuitively, this means that there are 
roughly $6.6 \times 10^7$ monopoles per every unit 
volume of the earth. This is a significant number, 
which suggests that there could be enough electroweak 
monopoles left over in the universe which we could 
detect.   
 
\section{Parker Bound on Monopole Density}

Obviously the monopoles are accelerated in magnetic 
fields, in particular the intergalactic magnetic field. 
It is well known that the average strength of the galactic 
magnetic field $B$ is about $3\mu G\simeq 1.2\times 
10^{-9} ~\text{T}$.  The energy gained by the monopole 
of charge $g_m =4\pi/e$ passing across the magnetic 
field $B$ of scale $L$ is 
\begin{gather}
\Delta E = \frac{4\pi}{e} L B  \simeq 1.2 \times 10^{11} 
\times \Big(\frac{L}{10^{21}~{\rm cm}} \Big)~\text{GeV}, 
\end{gather}
where $L$ is normalized to the typical coherence length 
of galactic magnetic field $L_0 =300~\text{pc}\simeq 
10^{21}~\text{cm}$. Traveling through the distance 
$L_0$, the monopole drains energy $\Delta E\simeq 
10^{11}~\text{GeV}$ from the magnetic field and becomes 
ultra-relativistic \cite{pin}. So, although the monopoles 
when decoupled around $29.5~\text{MeV}$ were 
non-relativistic, the remnant monopoles at present 
universe should be treated as relativistic.

Requiring that the rate of this energy loss in the galaxy 
is small compared to the time scale on which the galactic 
magnetic field can be regenerated, we can obtain 
the upper bound on the flux of the monopoles (with 
mass $M_m \le 10^{17}~\text{GeV}$),
\begin{gather}
F \le 10^{-15}~\text{cm}^{-2}\text{sec}^{-1}\text{sr}^{-1}.
\label{Pbound}
\end{gather}
This is the Parker bound \cite{parker}. 

The Parker bound sets a limit on the monopole density 
in the universe. The monopoles with velocity $v_m$ 
uniformly distributed throughout the universe generate 
the monopole flux \cite{turner}
\begin{gather}
F \simeq \frac{v_m}{4\pi }n_{m,0}
=\frac{v_m}{4\pi}\Big(\frac{\rho_{c,0}}{h^2 M_m} \Big)
\Omega_{mo} h^2   \nn\\
\simeq 2.3 \times 10^{-2}~\Omega_{mo} h^2
~\Big(\frac{v_m}{10^{-3}c}\Big) \nn\\
\times \Big(\frac{1~{\rm TeV}}{M_m}\Big)  
~\text{cm}^{-2}\text{sec}^{-1}\text{sr}^{-1},
\end{gather}
where $v_m$ is expressed in terms of the average virial 
velocity of the galaxy $10^{-3} c$. This, with (\ref{Pbound}), 
requires
\begin{gather}
\Omega_{mo} h^2 \lesssim 4.3\times 10^{-14}
\Big(\frac{10^{-3}c}{v_m} \Big)
\times \Big(\frac{M_m}{1~{\rm TeV}}\Big).
\label{Plimit}
\end{gather}
This set a stringent limit on the density for the relativistic 
($v_m \simeq c=1$) electroweak monopoles. This causes 
us a serious trouble, because our estimate of the density 
parameter (\ref{dparam0}) is roughly $10^{5}$ times bigger 
than this limit.  

We could think of possible ways to circumvent this trouble. 
First of all, we might suppose that the limit (\ref{Plimit}) 
is not trustable. Obviously this is an approximation. For 
example, the monopoles in the galactic magnetic field,
when accelerated near the velocity of light, will make 
the Chrenkov radiation and will slow down to a limiting 
velocity considerably less than the velocity of light.
  
But a more realistic explanation could be that most of 
the electroweak monopoles in the universe are actually 
buried inside the galactic cores. This is quite possible, 
because the electroweak monopoles become the natural 
source for the premodial black holes and the structure 
formation. Certainly, as the heaviest stable particles, 
they could easily cause the density pertubation and 
become the seeds for the large scale structures in 
the universe.  

Another possibility is that many of the electroweak 
monopoles are captured inside stellar objects when 
they hit (large) stellar objects, because they have 
a large capture cross section due to the magnetic 
interaction. In fact a relativistic electroweak monopole 
with mass 5 TeV is expected to travel less than 10 m 
in Aluminum before they are trapped \cite{bolo}.  
So the electroweak monopoles coming to the earth 
loose most of the energy passing through the earth 
atmospheric sphere and become non-relativistic, 
and could easily be trapped near the earth surface. 
In fact we may conjecture that these trapped monopoles 
(and buried in large scale structures) could have been 
the very source of the intergalactic magnetic field. 

More importantly, this strongly suggests that the stellar 
objects could have filtered out and diluted the density 
of the monopoles in the universe greatly, so that 
the remnant monopole density at present universe 
might have become much smaller than (\ref{C1bound}). 
So at this moment it is not clear whether our result 
is in contradiction with the Parker bound.   

\section{Discussions}   

In this paper we have studied the cosmic production 
and evolution of the electroweak monopole, and 
estimated the remnant monopole density in the present 
universe. Our analysis confirms that, although 
the electroweak phase transition is of the first order, 
it is very mildly first order. So the monopole
production mechanism is not the vacuum bubble 
collisions during the phase transition but the thermal 
fluctuations of the Higgs field after the phase transition,
which plant the seed of the monopoles. 

Our result shows that the electroweak monopoles are
produced when the temperature of the universe was 
around $100~{\rm GeV}$ (or about $10^{-11}~{\rm sec}$ 
after the big bang). And initially the monopoles are 
produced copiously, perhaps a bit too copiously to be 
realistic. But most of them are annihilated as soon as 
created, and this annihilation continues for a long time 
until the universe cools down to around 39.5 MeV, even 
after the muon pair annihilation. And eventually they 
decouple from the other matters at around $0.5~\text{MeV}$,
when the electron pairs annihilate and the monopole 
interaction rate becomes less than the Hubble expansion 
rate.  

Because of this the electroweak monopole density become 
very small, $\Omega_{mo} \simeq 6.3 \times 10^{-11}$, 
in the present universe. This tells that, unlike the grand 
unification monopole, the electroweak monopole can 
not overclose the universe. As importantly, this assures 
that the production of the electroweak monopole does 
not alter the standard cosmology in any significant way, 
and exclude the possibility the electroweak monopole 
to be a candidate of the dark matter.  

On the other hand, this means that there are enough 
monopoles left over, roughly $6.6 \times 10^7$ monopoles 
per every unit volume of the earth in the present universe. 
This strongly indicates that experimentally there are 
enough remnant monopoles that we should be able to 
detect without much difficulty. 

However, we should to keep in mind the possibility 
that the actual density of the remnant monopoles 
could be much less than this. This is because many 
of them could have been buried in the the large 
scale structures of the universe and/or filtered out 
by stellar objects. In fact, the actual monopole density 
could be smaller by the factor $10^{-5}$, or about 
$10^3$ per every unit volume of the earth, as the Parker 
bound indicates.  

Our result could provide useful informations for 
the monopole detection experiments. The recently 
upgraded 13 TeV LHC at CERN might have finally 
reached the threshold energy to produce just one 
electroweak monopole-antimonopole pair. In this 
case MoEDAL has a best chance to detect the monopole. 
However, it is not clear if LHC could actually produce 
the monopole pair. This is an important issue, because 
if the monopole mass is larger than 6.5 TeV, LHC is 
not supposed to be able to produce the monopole. 

Another issue is the monopole production mechanism 
at LHC. It has generally been believed that LHC could 
produce the monopole pair by Drell-Yan process 
($p\bar p \rightarrow \gamma \rightarrow M \bar M$)
and/or photon fusion process ($p\bar p \rightarrow 
p \bar p +\gamma \gamma \rightarrow p \bar p 
+M \bar M$)  \cite{dypf}. This would be the case if we 
treat the monopole as a point particle. But since 
the monopole is a topological particle, it is not clear 
if this is correct. Even if this picture is correct, LHC
certainly need the change of topology to create 
the monopole pair, and we have to explain how LHC 
can achieve this during the collision. 

Our analysis suggests that the cosimc monopole production 
mechanism should apply equally well to LHC. In other words, 
LHC could produce the monopole pair when the colliding 
beam core cools down from the maximum temperature 
13 TeV in the symmetric phase to the electroweak phase 
around 100 GeV. At this temperature the thermal fluctuation 
of the Higgs field makes the monopole seeds, the baby 
monopoles, of mass around 1 TeV. This is several times 
less than the monopole mass at zero temperature which 
is expected to be several TeV. This strongly implies that 
the energy constraint of the LHC on the production of 
the electroweak monopole may not be so strong obstacle 
as we have thought. 

To produce the monopole-antimonopole pair, however, 
our analysis tells that LHC should satisfy two more 
conditions. First, it should generate the hot plasma of 
the beam core of the size at least as big as the initial 
correlation length $\xi_i\simeq 10^{-15}~\text{cm}$. 
Second, this beam core should last at least ten times 
longer than $10^{-26}~\text{sec}$ to allow enough 
fluctuations for the Higgs vacuum to plant the monopole 
seed. 

Fortunately, the size of the beam core at LHC has 
been measured to be about $0.03~\text{cm}$ and 
lasts for about $10^{-12}~\text{sec}$, long enough 
for the Higgs field to allow more than enough thermal 
flucuations \cite{lhc}. This strongly implies that there 
is a good chance that LHC could actually produce one 
monopole-antimonopole pair, which MoEDAL could
detect. 

For the other experiments searching for the remnant 
monopoles in the universe, in particular for IceCube, 
ANTARES, and Auger, an important thing is to know 
the characteristic features of the remnant 
monopoles \cite{icecube,antares,auger}. Although 
the electroweak monopoles were non-relativistic 
when they decoupled, the intergalactic magnetic field 
makes them highly relativistic at present universe. 
On the other hand, they loose most of the energy 
passing through the earth atmospheric sphere, so that 
near the earth surface they become non-relativistic 
again. So on earth these experiments should look 
for the non-relativistic monopole coming from 
the sky, which has mass around 4 to 10 TeV 
and magnetic charge $4\pi/e$.  

In principle these experiments have the ability to detect 
the monopoles, but there is one catch here. These 
detectors (except Auger) are located underground. 
This could cause a problem, because the non-relativistic 
remnant monopoles may be trapped before they reach 
the detector. For example, IceCube is located under 
the 2 km thick iceberg, which the remnant monopoles 
may not be able to penetrate. If so, it would be very 
difficult for IceCube to detect the monopole. In fact 
this could be a main reason why IceCube could not 
find it. This suggests that a best way to detect the remnant 
monopole is to install the detector at high altitude. So 
these experiments have to find a way to circumbent 
this problem to detect the remnant monopoles.
 
Although the remnant electroweak monopoles in 
the present universe are insignificant, they have 
important physical implications. As we have speculated, 
they could have been the seed of the large scale 
structures in the universe. Indeed the electroweak 
monopoles with mass about $10^4$ times heavier 
than the proton, could easily generate the density 
perturbation and become an excellent candidate 
for the seed of the large scale structures in the universe. 
Moreover, as the heaviest relativistic magnetically 
charged particles in the universe, they become 
the source of ultra-high energy cosmic rays. 
Furthermore, they could play an important role in 
the electroweak baryogenesis. Clearly this is a very 
interesting possibility need to be studied further. 

But the most important point of the electroweak 
monopole is that it must exist. This makes the detection 
of the electroweak monopole the final (topological) 
test of the standard model. In spite of huge efforts,
however, the monopole detection so far has not 
been successful. There could be two reasons for this. 
First, many of these experiments were the blind 
seraches in the dark room, with few theoretical 
leads. Second, many were looking for different type 
of monopoles, in particular the grand unification 
monopole. 

So, aiming at the electroweak monopole which has 
the unique characteristic features, we could enhance 
the probability to detect the monopole greatly. We 
hope that our analysis in this paper could help  
confirm the existence of the electroweak monopole.

{\bf ACKNOWLEDGMENT}

~~~The authors thank James Pinfold for the careful reading 
of the manuscript and valuable advice to improve 
the paper. The work is supported in part by the National 
Research Foundation of Korea funded by the Ministry of 
Education (Grants 2015-R1D1A1A0-1057578 and 
2015-R1D1A1A0-1059407), and by Konkuk University.


\begin{thebibliography}{99}
\bibitem{dirac} P.A.M. Dirac, Proc. Roy. Soc. London, 
{\bf A133}, 60 (1931); Phys. Rev. {\bf 74}, 817 (1948).
\bibitem{wu} T.T. Wu and C.N. Yang, in {\it Properties 
of Matter under Unusual Conditions}, edited by 
H. Mark and S. Fernbach (Interscience, New York) 1969; 
Phys. Rev. {\bf D12}, 3845 (1975);
Y.M. Cho, Phys. Rev. Lett. {\bf 44}, 1115 (1980); 
Phys. Lett. {\bf B115}, 125 (1982).
\bibitem{thooft} G. 't Hooft, Nucl. Phys. {\bf B79}, 
276 (1974); A.M. Polyakov, JETP Lett. {\bf 20}, 194 (1974); 
M. Prasad and C. Sommerfield, Phys. Rev. Lett. {\bf 35}, 
760 (1975).
\bibitem{dokos} C. Dokos and T. Tomaras, Phys. Rev. {\bf D21}, 
2940 (1980).
\bibitem{plb97} Y.M. Cho and D. Maison, Phys. Lett. {\bf B391}, 
360 (1997).
\bibitem{yang} Yisong Yang, Proc. Roy. Soc. London, {\bf A454}, 
155 (1998); Yisong Yang, {\it Solitons in Field Theory and 
Nonlinear Analysis} (Springer Monographs in Mathematics), 
p. 322 (Springer-Verlag) 2001.
\bibitem{epjc15} Kyoungtae Kimm, J.H. Yoon, and Y.M. Cho, 
Eur. Phys. J. {\bf C75}, 67 (2015); Kyoungtae Kimm, 
J.H. Yoon, S.H. Oh, and Y.M. Cho, Mod. Phys. Lett. 
{\bf A31}, 1650053 (2016).
\bibitem{ellis} J. Ellis, N.E. Mavromatos, and T. You, 
Phys. Lett {\bf B756}, 29, (2016).
\bibitem{plb16} Y.M. Cho, Kyoungtae Kimm, J.H. Yoon, 
Phys. Lett. {\bf B761}, 203 (2016).

\bibitem{medal1} B. Acharya et al. (MoEDAL Collaboration),
Phys. Rev. Lett. {\bf118}, 061801 (2017).
\bibitem{medal2} B. Acharya et al. (MoEDAL Collaboration),
JHEP 1608, 067 (2016); Int. J. Mod. Phys. {\bf A29}, 
1430050 (2014); Y.M. Cho and J. Pinfold, Snowmass white 
paper, arXiv: hep-ph/1307.8390.
\bibitem{icecube} R. Abbasi et al. (IceCube Collaboration) 
Phys. Rev. {\bf D87}, 022001 (2013); M. Aartsen et al. 
(IceCube Collaboration), Eur. Phys. J. {\bf C74}, 2938 (2014).
\bibitem{antares} S. Adri{\'a}n-Martinez et al. (ANTARES 
Collaboration), Astropart. Phys. {\bf 35}, 634 (2012).
\bibitem{auger} A. Aab et al. (Pierre Auger Collaboration)
Phys. Rev. {\bf D94}, 082002 (2016).

\bibitem{kibble} T.W.B. Kibble, J. Phys. {\bf A9} 1387 (1976).
\bibitem{pres} J.P. Preskill, Phys. Rev. Lett. {\bf 43}, 1365 (1979). 
\bibitem{guth} A. Guth and E. Weinberg, Nucl. Phys. 
{\bf B212}, 321 (1983).
\bibitem{zurek}  W.H. Zurek, Phys. Rep. {\bf 276} 177 (1996).
\bibitem{infl} A. Guth, Phys. Rev. {\bf D23}, 347 (1981);
A. Linde, Phys. Lett. {\bf B108} 389 (1982).

\bibitem{kriz} D. Krizhnits and A. Linde, Phys. Lett. {\bf B42}, 
471 (1972); C. Bernard, Phys. Rev. {\bf D9}, 3312 (1974);
L. Dolan and R. Jakiw, Phys. Rev. {\bf D9}, 3320 (1974);
S. Weinberg, Phys. Rev. {\bf D9}, 3357 (1974).
\bibitem{zel} Y.B. Zel'dovich and M.Yu. Khlopov, Phys. Lett. 
{\bf 79B}, 239 (1978); P. Adams,V. Canuto, and H.Y. Chiu, 
Phys. Lett. {\bf 61B}, 397 (1976).
\bibitem{and} G. Anderson and L. Hall, Phys. Rev. {\bf D45}, 
2685 (1992); M. Dine, R. Leigh, P Huet, A. Linde, and D. Linde, 
Phys. Rev. {\bf D46}, 550 (1992).
\bibitem{koba} S. Arunasalam and A. Kobahhidze,
arXiv: hep-ph/1702.04068 (2017).

\bibitem{nogo} S. Coleman, {\it Aspects of Symmetry} (Cambridge 
Univ. Press, 1985); T. Vachaspati and M. Barriola, Phys. Rev. 
Lett. {\bf 69}, 1867 (1992).

\bibitem{vil} A. Vilenkin and E. Shellard,  
{\it Cosmic Strings and other Topological Defects} 
(Cambridge University Press) 1994.
\bibitem{pin} S. Burdin, M. Fairbaim, P. Mermod, D. Milstead,
J. Pinfold, T. Sloan, and W. Taylor, Phys. Rep. {\bf 582}, 1 (2015).

\bibitem{parker} E.N. Parker, Astrophys. J. 160, 383 (1970).
\bibitem{turner} M.S. Turner, E.N. Parker, and T.J. Bogdan, 
Phys. Rev. {\bf D26}, 1296 (1982); F.C. Adams, M. Fatuzzo, 
K. Freese, G. Tarl\'e, R. Watkins, and M.S. Turner, Phys. Rev. Lett. 
{\bf 70}, 2511(1993).
\bibitem{bolo} S. Cecchini, L. Patrizii, Z. Sahnoun, G. Sirri, 
and V. Togo, arXiv: hep-ph/1606.01220.
\bibitem{dypf} T. Dougall and S. Wick, Euro. Phys. J. 
{\bf A39}, 213 (2009); L. Epele, H. Fanchiotti, C. Cannl, V. Mitsou, 
and V. Vento, Euro. Phys. J. Plus {\bf 127}, 60 (2012).
\bibitem{lhc} Redaelli, Stefano, et al., FERMILAB-CONF-15-135-AD-APC, 
2015.

\end{thebibliography}
\end{document}